\documentclass[10pt, a4paper, twocolumn]{article}

\usepackage{authblk}
\usepackage{geometry}
\usepackage{mathpazo}
\usepackage{cite}
\usepackage{amsmath,amssymb,amsfonts}
\usepackage{graphicx}

\usepackage{tikz}
\usetikzlibrary{positioning}
\tikzset{
    mycircle/.style={circle, draw, minimum size=1.0cm, inner sep=0pt, text=white, align=center},
    myedge/.style={midway, sloped, above},
    every path/.style={draw, thick, -latex}
}
\definecolor{cmblue}{HTML}{1f77b4}
\definecolor{cmorange}{HTML}{ff7f0e}
\definecolor{cmgreen}{HTML}{2ca02c}
\definecolor{cmred}{HTML}{d62728}
\definecolor{cmpurple}{HTML}{9467bd}
\definecolor{cmbrown}{HTML}{8c564b}
\definecolor{cmpink}{HTML}{e377c2}
\definecolor{cmgray}{HTML}{7f7f7f}
\definecolor{cmyellow}{HTML}{bcbd22}
\definecolor{cmcyan}{HTML}{17becf}
\newcommand\checkmarks[1][]{%
  \tikz[scale=0.4,#1]{\fill(0,.35) -- (.25,0) -- (1,.7) -- (.25,.15) -- cycle;}%
}
\newcommand\crossmark[1][]{%
  \tikz[scale=0.4,#1]{
    \fill(0,0)--(0.1,0) .. controls (0.5,0.4) .. (1,0.7)--(0.9,0.7) ..  controls (0.5,0.5) ..(0,0.1) --cycle;
    \fill(1,0.1)--(0.9,0.1) .. controls (0.5,0.3) .. (0,0.7)--(0.1,0.7) .. controls (0.5,0.4) ..(1,0.2) --cycle;
  }%
}

\usepackage{listings}
\lstdefinestyle{bash}{
  language=bash,
  basicstyle=\ttfamily\footnotesize,
  keywordstyle=\color{black},        %
  commentstyle=\color{gray},         %
  stringstyle=\color{red},           %
  numbers=left,                      %
  numberstyle=\scriptsize\color{gray}, %
  stepnumber=1,                      %
  xleftmargin=2em,
  breaklines=true,                    %
  frame=tb,                           %
  captionpos=b,                       %
  emph={ethtool,tc},                  %
  emphstyle=\color{blue}              %
}

\usepackage[table]{xcolor}
\usepackage{booktabs}
\usepackage{nicefrac}
\usepackage{tcolorbox}
\usepackage{tabularx}
\usepackage[hyphens]{url}

\newcommand{\ms}[1]{%
    \relax\ifmmode
        \mathord{\mathcode`\-="702D\it #1\mathcode`\-="2200}%
    \else
        {\it #1}%
    \fi
}

\newcommand{\code}[1]{{\fontfamily{cmss}\selectfont\textcolor{gray}{#1}}}

\newcommand{\lit}[1]{%
    \relax\ifmmode
        \mathord{\mathcode`\-="702D\sf #1\mathcode`\-="2200}%
    \else
        {\it #1}%
    \fi
}

\usepackage{xspace}
\newcommand{\quic}{\textsc{Quic}\xspace}

\usepackage{flushend}

\usepackage{subcaption}

\usepackage{varioref}
\usepackage[hidelinks]{hyperref}
\usepackage[capitalize]{cleveref}

\begin{document}

\title{Blockchain Communication Vulnerabilities}

\author[1]{Andrei Lebedev}
\author[1,2]{Vincent Gramoli}
\affil[1]{The University of Sydney}
\affil[2]{Redbelly Network}
\date{}

\maketitle

\begin{abstract}
    Blockchains are diverse in the way they handle communications between their nodes to disseminate information, mitigate attacks, and agree on the next block. While security vulnerabilities have been identified, they rely on an attack custom-made for a specific blockchain communication protocol. 
    To our knowledge, the vulnerabilities of multiple blockchain communication protocols to adversarial conditions have never been compared.

    In this paper, we compare empirically the vulnerabilities of the communication protocols of five modern in-production blockchains, Algorand, Aptos, Avalanche, Redbelly and Solana, when attacked in five different ways. 
    We conclude that Algorand is vulnerable to packet loss attacks, Aptos is vulnerable to targeted load attacks and leader isolation attacks,
    Avalanche is vulnerable to transient failure attacks, Redbelly's performance is impacted by  packet loss attacks and Solana is vulnerable to stopping attacks and leader isolation attacks.
    Our system is open source.
\end{abstract}

\section{Introduction}\label{sec:introduction}

Blockchains have been known to be vulnerable to network attacks. Their, often implicit~\cite{But17}, synchrony assumption~\cite{DLS88} states that all messages have to be delivered in a known bounded amount of time for them to work. 
Because this assumption is unrealistic in an open network like the Internet, various hacks happened:
An ISP hack led to the theft of funds in Bitcoin~\cite{Guar14}, the balance attacks against Ethereum~\cite{NG17,NTT21,NEJ24} are all based on introducing network delays, either through man-in-the-middle~\cite{EGJ18} or BGP-hijacking~\cite{AZV17}. In the past few years, Bitcoin lost $\mathdollar 70,000$~\cite{Red20} while Ethereum Classic lost $\mathdollar 5.6$ million~\cite{Zim20}. These attacks demonstrate that communication protocols raise risks in classic blockchains.
Unfortunately, as of today the network protocols of modern blockchain technologies have never been compared. 

In this paper, we compare the security of modern blockchain communication protocols.
We disregard Ethereum and Bitcoin that are known to suffer from a lack of synchrony and instead focus our comparison on the following five modern blockchains: 
\textit{(1)}~Algorand is claimed to not violate safety in a partially synchronous environment~\cite{GHM17},
\textit{(2)}~Avalanche was proven to work in a synchronous environment and the proof is claimed generalizable without full synchrony~\cite{rocket_scalable_2020}, 
\textit{(3)}~Aptos~\cite{noauthor_aptos_2022} and \textit{(4)}~Redbelly~\cite{CNG21} are blockchains supposed to cope with unexpected message delays, 
\textit{(5)}~Solana~\cite{Sol25} is a recent blockchain that became popular for launching meme coins, including those from the U.S. president's family, it uses erasure coding to cope with packet losses.
These blockchains offer diverse communication protocols. The communication protocols of
Algorand and Aptos
aim at hiding validator nodes, responsible for deciding upon the next block, by placing them behind intermediary nodes~\cite{CGT19} and validator full nodes~\cite{apt-nodes}, respectively.
The communication protocols of Avalanche and Redbelly expose validator nodes to Denial-of-Service (DoS) attempts but protect them with throttling and rate limiting, respectively, dropping selected requests that would otherwise consume too many resources. 
The communication protocol of Solana exploits a hierarchical overlay in order to maximize dissemination~\cite{noauthor_solana_nodate} at times preventing direct access to validator nodes~\cite{noauthor_aptos_2022}.

In order to compare the security of blockchain communication protocols on the same ground, 
we surveyed attacks tailored to specific blockchains~\cite{Ros12,HKZG15,WG16,NG17,ES18,STM18,MJP20,MMM20,AFO21,NTT21,IST21,CBR22,NEJ24,SPL25} and derived these five new attacks 
impairing communication protocols in a blockchain-agnostic way:
\begin{description}
    \item[Targeted load attack.] An attack, inspired by DoS attacks~\cite{STM18,MJP20,CBR22}, that sends traffic to one blockchain node.
    \item[Transient failure attack.] An attack, inspired by the continuous churn~\cite{MMM20,IST21}, that affects few nodes transiently.
    \item[Packet loss attack.] An attack, inspired by balance attacks~\cite{NG17,NTT21,NEJ24}, that drops some network packets.
    \item[Stopping attack.] An attack that crashes part of a blockchain network, inspired by majority attacks~\cite{Ros12,ES18,AFO21}.
    \item[Leader isolation attack.] \sloppy{An attack, inspired by eclipse attacks~\cite{HKZG15,WG16,SPL25}, that isolates one blockchain node.}
\end{description}

\begin{table*}
  \caption{
  Vulnerabilities of different blockchains under attacks, alongside claimed fault tolerance threshold reported by the blockchain team, expressed as a function of the number $n$ of blockchain nodes. A red cross \crossmark[red, scale=1] indicates significant vulnerability while a checkmark \checkmarks[blue] indicates resilience observed under the tested attack conditions.
  Dashes `{\bf ---}' indicate inconclusive or inapplicable scenarios. (*) Note that, for the probabilistic protocols Algorand and Avalanche, we selected the threshold of $\nicefrac{n}{5}$ from their original papers that lowers the probability of a global failure to below $10^{-9}$.}\label{table:comparison}
  \centering
  \small{
  \setlength{\tabcolsep}{1pt}
  \begin{tabular*}{\textwidth}{@{\hspace{\tabcolsep}\extracolsep{\fill}} lcccccccc @{\hspace{\tabcolsep}}}
    \toprule
    {\bf Blockchain} & {\bf Reference} & {\bf Communication} & {\bf Fault tolerance ($f$)} & {\bf Load} & {\bf Failure} & {\bf Loss} & {\bf Stopping} & {\bf Isolation} \\
    \midrule
    {\bf Algorand}  & \cite{GHM17} & Gossip & \color{red}{$\nicefrac{n}{5}$}* & \checkmarks[blue]         & \checkmarks[blue]  & \crossmark[red, scale=1]          & \checkmarks[blue] & \checkmarks[blue]       \\
    {\bf Aptos}  & \cite{noauthor_aptos_2022}  & Hierarchical & \color{blue}{$\nicefrac{n}{3}$} & \crossmark[red, scale=1]         & {\bf ---}       & {\bf ---}          & \checkmarks[blue] & \crossmark[red, scale=1]       \\
    {\bf Avalanche}  & \cite{rocket_scalable_2020} & Throttled & \color{red}{$\nicefrac{n}{5}$}* & {\bf ---}       & \crossmark[red, scale=1]          & {\bf ---}          & {\bf ---} &  {\bf ---}     \\
    {\bf Redbelly}   & \cite{CNG21} & Rate limited & \color{blue}{$\nicefrac{n}{3}$} & \checkmarks[blue]         & \checkmarks[blue]      & {\bf ---}             & \checkmarks[blue] & \checkmarks[blue]         \\
    {\bf Solana}     & \cite{Yak21} & Hierarchical & \color{blue}{$\nicefrac{n}{3}$} & \checkmarks[blue]          & {\bf ---}        & \checkmarks[blue]           & \crossmark[red, scale=1] & \crossmark[red, scale=1]        \\
    \bottomrule
  \end{tabular*}}
\end{table*}

The results of our study, summarized in \cref{table:comparison} but detailed in later sections, lead to four interesting conclusions.

\begin{description}

\item[Crashing blockchains.] We noticed that some blockchains, like Avalanche and Solana, stop servicing requests under some attacks. First, Avalanche is vulnerable to transient failure attacks because the combination of its throttling with transient failures prevents it from committing transactions.
Second, Solana is vulnerable to stopping attacks in that if sufficiently many nodes experience a transient outage, then the system may get stuck even after these nodes recover.

\item[Transport protocol impact.] The selected network transport protocol is instrumental in the quality of information dissemination. Redbelly's performance is particularly impacted by packet loss attacks due to its TCP-based protocol that prevents it from recovering lost transactions. This is generally common to all TCP-based blockchains, including Algorand, Aptos and Avalanche. But Solana disseminates blocks successfully even under very adversarial packet losses due to its combination of \quic and erasure coding.

\item[Single node isolation.] Aptos and Solana are vulnerable to the leader isolation attacks as Aptos reputation-based leader selection and Solana leader schedule are deterministic and allow attackers to predict and target successive leaders. Targeting a single node of the network can thus translate into a global slowdown of these blockchain networks.

\item[Node complexity exposure.] \sloppy{Some blockchains hide CPU-intensive tasks that can be triggered with minimal effort from the attacker. In particular, Aptos is vulnerable to targeted load attacks as a single validator can become overloaded by the signatures it has to collect when trying to avoid producing a quadratic number of signatures.}

\end{description}

\Cref{sec:background} presents the background, 
\cref{sec:settings} presents the experimental settings to compare the security of blockchains.
\Cref{sec:workload} presents the problem of responding to requests under load saturation attacks. 
\Cref{sec:stability} assesses vulnerability of blockchains to transient failure attacks, 
\cref{sec:messageloss} evaluates their resistance to packet loss attacks, 
\cref{sec:liveness} examines their vulnerability to stopping attacks and \cref{sec:isolation} analyzes impact of leader isolation attacks on them.
 \Cref{sec:discussion} discusses two critical issues present in Avalanche and Solana and their countermeasures, \cref{sec:relatedwork} presents the related work and \cref{sec:conclusion} concludes.
 
\begin{tcolorbox}[title=Responsible Disclosure, colback=gray!5!white,colframe=gray!75!gray,boxsep=2pt,left=2pt,right=2pt,top=2pt,bottom=2pt]
In accordance with responsible disclosure principles, we have notified the respective development and security teams of the vulnerabilities and findings discussed in this paper prior to publication. This includes teams associated with Algorand, Aptos, Avalanche (Ava Labs), Solana (Anza), and Redbelly, providing them the opportunity to investigate and review our results. As a result,
Anza confirmed that Solana experiences vulnerabilities.
\end{tcolorbox}

\section{Background}\label{sec:background}

In this section, we review  
the communication protocols of the modern in-production blockchains (cf.~\cref{table:comparison}) supposedly secure that we evaluate here.

\subsection{The Algorand communication protocol}\label{sec:algorand}

\begin{figure*}
    \centering
    \begin{subfigure}[b]{0.2\textwidth}
        \centering
        \begin{tikzpicture}[
            mynode/.style={mycircle, fill=cmblue},
            node distance=8mm
        ]
        \node[mynode] (A) {non\\relay};
        \node[mynode, below=of A] (B) {relay};
        \node[mynode, right=of A] (C) {non\\relay};
        \node[mynode, below=of C] (D) {relay};

        \draw[<->] (A) -- (B) node[myedge] {cons};
        \draw[<->] (C) -- (D) node[myedge] {sync};
        \draw[<->] (B) -- (C) node[myedge] {cons};
        \end{tikzpicture}
        \caption{Algorand\label{sfig:algorand}}
    \end{subfigure}%
    \begin{subfigure}[b]{0.2\textwidth}
        \centering
        \begin{tikzpicture}[
            mynode/.style={mycircle, fill=cmorange},
            node distance=8mm
        ]

        \node[mynode] (A) {valid.};
        \node[mynode, below=of A] (B) {valid.};
        \node[mynode, left=of A] (C) {VF};
        \node[mynode, below=of C] (E) {PubF};

        \draw[<->] (A) -- (B) node[myedge] {cons};
        \draw[<->] (A) -- (C) node[myedge] {sync};
        \draw[<->] (C) -- (E) node[myedge] {req};
        \end{tikzpicture}
        \caption{Aptos\label{sfig:aptos}}
    \end{subfigure}%
    \begin{subfigure}[b]{0.2\textwidth}
        \centering
        \begin{tikzpicture}[
            mynode/.style={mycircle, fill=cmgreen},
            node distance=8mm
        ]
        
        \node[mynode] (A) {valid.};
        \node[mynode, right=of A] (B) {valid.};
        \node[mynode, below=of A] (C) {PF};
        \node[mynode, below=of B] (D) {AF};
        
        \draw[<->] (A) -- (B) node[myedge] {cons};
        \draw[<->] (A) -- (C) node[myedge] {sync};
        \draw[<->] (D) -- (B) node[myedge] {sync};
        
        \end{tikzpicture}
        \caption{Avalanche\label{sfig:avalanche}}
    \end{subfigure}%
    \begin{subfigure}[b]{0.2\textwidth}
        \centering
        \begin{tikzpicture}[
            mynode/.style={mycircle, fill=cmred},
            node distance=8mm
        ]
        
        \node[mynode] (A) {gov.};
        \node[mynode, right=of A] (B) {gov.};
        \node[mynode, below=of A] (C) {cand.};
        \node[mynode, below=of B] (D) {cand.};
        
        \draw[<->] (A) -- (B) node[myedge] {cons};
        \draw[<->] (A) -- (C) node[myedge] {sync};
        \draw[<->] (B) -- (D) node[myedge] {sync};
        \draw[<->] (C) -- (B);
        \draw[<->] (D) -- (A) node[pos=0.5, fill=white, inner sep=2pt] {sync}; 
        
        \end{tikzpicture}
        \caption{Redbelly\label{sfig:redbelly}}
    \end{subfigure}%
    \begin{subfigure}[b]{0.2\textwidth}
        \centering
        \begin{tikzpicture}[
            mynode/.style={mycircle, fill=cmpurple}, 
            node distance=8mm 
        ]
        
        \node[mynode] (root) {root};
        \node[mynode, right=of root] (leader) {leader};
        \node[mynode, below=of root] (valid2) {valid.};
        \node[mynode, below=of leader] (valid1) {valid.};
        
        \draw[<->] (root) -- (leader) node[myedge] {sync};
        \draw[<->] (root) -- (valid1) node[myedge] {sync};
        \draw[<->] (root) -- (valid2) node[myedge] {sync};
        
        \end{tikzpicture}
        \caption{Solana\label{sfig:solana}}
    \end{subfigure}%
    \caption{The different topologies used by the different blockchains where valid., VF, PubF, AF, PF, gov. and cand. stand for validator, validator full, public full, archival full, pruned full, governor and candidate, respectively, and where cons, sync and req stand for consensus, synchronization and request.}
    \label{fig:enter-label}
\end{figure*}
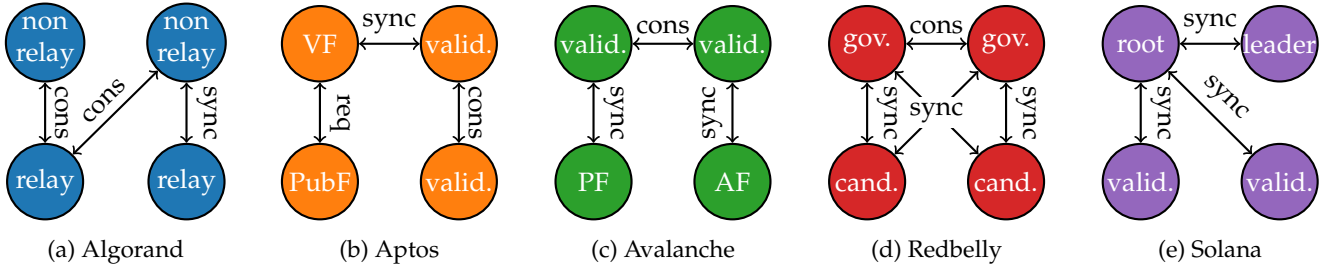

Algorand~\cite{GHM17} is a blockchain protocol that shuffles consensus participants via %
Verifiable Random Functions (VRFs). 
Consensus participants then run the $\ms{BA\star}$ protocol to reach an agreement on a block despite the presence of Byzantine faults. 
Each node validates each message before sending it at most once to each other node~\cite{CGT19} through gossip. 
To this end, each node maintains one TCP connection per node in its neighborhood, which offers WebSockets over HTTP. Different nodes can play different roles as depicted in \vref{sfig:algorand}, where relay nodes connect to (potentially multiple) non-relay nodes that run consensus but without being connected to each other. These nodes can be archival and participating nodes~\cite{algo-node-types}.

\subsection{The Aptos communication protocol}\label{sec:aptos}

Aptos~\cite{noauthor_aptos_2022} is a leader-based blockchain that builds upon a variant of PBFT~\cite{castro_practical_1999} with a 
quadratic communication complexity view-change and a cubic communication complexity.
To cope with the leader bottleneck~\cite{VG20} and redundant validations~\cite{THG23}, Aptos features a Quorum Store optimization.
Validators
repeat the following steps in parallel: \textit{(1)}~Pull transactions from the mempool; \textit{(2)}~Arrange transactions into batches; 
\textit{(3)}~Broadcast batches to all other validators; \textit{(4)}~Persist received batches, sign their digests, and send back signatures; and \textit{(5)}~Collect signatures from more than $\nicefrac{2n}{3}$ nodes to form a \emph{proof-of-store}.
It allows validators to asynchronously broadcast transactions, offloading the leader's network interface during the consensus protocol execution~\cite{apt-quorum-store}.
The communication model of Aptos builds upon TCP and is hierarchical in that nodes play different roles at different levels~\cite{apt-nodes}. As indicated in \cref{sfig:aptos}, the validator nodes are at the top level and run the consensus, their children are validator full (VF) nodes that download the current states from the validator nodes. The leaves are the public full (PubF) nodes that download the blocks from the VF nodes but cannot communicate directly with the top level (i.e., validator nodes).

\subsection{The Avalanche communication protocol}\label{ssec:avalanche}

Avalanche is probabilistic and
converges
towards an agreement using the Snowflake binary consensus protocol~\cite{rocket_scalable_2020}.
Nodes communicate over TCP and exploit throttling~\cite{noauthor_avalanchego_nodate} 
to cap the amount of CPU, disk, bandwidth, and message handling other nodes consume.
As depicted in \vref{sfig:avalanche}, Avalanche features archive full (AF) nodes and pruned full (PF) nodes that synchronize with the validators that, in turn, run the consensus~\cite{ava-nodes}.
Upgrades introducing dynamic fees and incremental increases in gas targets 
are discussed later in~\cref{sec:avalanche-target-increase}.
A dynamic proposer selection algorithm manages network load and block production.
Based on a seed derived from the parent block's height and chain ID, it 
pseudo-randomly selects a sequence of potential proposers for the next block height.
This sequence indicates in which order proposers append and sign new blocks.

\subsection{The Redbelly communication protocol}
Redbelly Blockchain~\cite{CNG21} is a scalable blockchain built on the DBFT consensus algorithm~\cite{CGLR18} that is \emph{leaderless} (i.e., non leader-based) and does not require synchrony.
DBFT has been formally verified with parameterized model checking~\cite{BGL22}, showing that it solves the consensus problem in all %
executions and for any system size.
A collaborative approach guarantees that superblocks with as many valid proposed blocks as possible~\cite{CNG21} are appended. 
Nodes communicate using TCP. As depicted in \cref{sfig:redbelly}, 
among all nodes, called candidates, a random subset are transiently promoted to governors and exchange messages directly while running a given consensus instance to mitigate resource waste~\cite{Gra22}.
Candidate nodes synchronize with multiple governors to cope (deterministically) with Byzantine behaviors. 
Each candidate, governor or client is rate limited to mitigate DoS attacks.
Redbelly  features the Scalable version of the Ethereum Virtual Machine (SEVM) and runs \emph{decentralized applications (dApps)}
written in Solidity~\cite{THG23}.

\subsection{The Solana communication protocol}\label{ssec:solana}

Solana~\cite{noauthor_solana_nodate} is a leader-based blockchain. Its transactions need 30 subsequent blocks (at least 12.4\,s) to be committed.
Its leaders are scheduled for four slots and each
schedules validators based on their stake to specific slots within an epoch. 
Validators vote on fork branches: a lockout period doubles with every successive vote on the same branch, so that a validator would wait for this lockout period to expire before voting on a conflicting branch.
This mechanism promotes commitment to the heaviest branch, though it can cause liveness stalls if many validators are simultaneously locked out and unable to vote on new slots.
Nodes communicate over the \quic network protocol~\cite{IETF21} to exchange transactions.
Nodes split blocks into chunks that they disseminate in a hierarchical structure, called Turbine~\cite{anza-turbine}, through UDP. %
As depicted in \cref{sfig:solana} the leader propagates a block by sending it to the root node that then disseminates it~\cite{helius-turbine}. Transactions are totally ordered with timestamps generated by a proof-of-history. %

\section{Comparing Blockchain Security}\label{sec:settings}

In this section, we 
list our assumptions, define the adversary and present our experimental setup.

\subsection{System model}%
We consider a blockchain network as a set of \emph{nodes}
committing transactions sent by \emph{clients}.
In order to model a realistic open network, where nodes can collude and impact the communication, we define an adversary that is either \emph{internal} and controls a subset of the nodes (e.g., a coalition) or \emph{external} and controls only the network (e.g., forging or dropping messages), similarly to Giuliari et al.~\cite{GSF24}.

\subsection{Threat model}%
In order to model a realistic open network, where nodes can collude and impact the communication, we define an adversary that is either \emph{internal} and controls a subset of the nodes (e.g., a coalition) or \emph{external} and controls only the network (e.g., forging or dropping messages), similarly to Giuliari et al.~\cite{GSF24}. More precisely, the adversary can execute the following five protocol-agnostic attacks, which form the basis of our ``adversarial communication setting'':
\textit{(i)}~Targeted load attack: an attack, inspired by denial-of-service (DoS) attacks~\cite{STM18,MJP20,CBR22}, that sends a constant rate of requests to a single node of its blockchain network; \textit{(ii)}~Transient failure attack: an attack, inspired by the continuous churn in blockchain networks~\cite{MMM20,IST21}, that crashes a relatively small portion of the blockchain network during a short period of time; \textit{(iii)}~Packet loss attack: an attack, inspired by balance attacks~\cite{NG17,NTT21,NEJ24}, that drops a portion of the packets exchanged between two parts of the blockchain network; \textit{(iv)}~Stopping attack: an attack that crashes a large portion of the blockchain network, inspired by majority attacks~\cite{Ros12,ES18,AFO21}, to stop the blockchain service; and \textit{(v)}~Leader isolation attack: \sloppy{an attack, inspired by eclipse attacks~\cite{HKZG15,WG16,SPL25}, that isolates the leader of the consensus protocol used by the blockchain service.}

\subsection{Experimental setup}
We deploy five types of blockchain networks: \textit{(i)}~Algorand v3.27.0 where nodes are relays, \textit{(ii)}~Aptos v1.25.1,  \textit{(iii)}~Avalanche C-Chain v1.12.1 and \textit{(iv)}~Solana Agave v2.0.20 where nodes are validators, and \textit{(v)}~Redbelly v0.36.2 where nodes are governors.
Each machine mimics the commodity computer run by an individual in a blockchain network, so, to obtain realistic results, we selected virtual machines (VMs) each with 4 vCPUs and 8\,GB of memory.
Note that this specification is lower than what some blockchains typically recommend, including 
Aptos~\cite{apt-reqs}, 
Avalanche~\cite{ava-fundamentals} or 
Redbelly~\cite{redbelly-reqs}, however, strict hardware requirements on remote nodes remain hard to enforce and a unique configuration is necessary for our comparison.
To assess security, a small network is sufficient. We thus
deploy distributed systems of 25 VMs running Ubuntu 24.04.1 LTS on top of a Proxmox cluster of physical servers, each equipped with 4x AMD Opteron 6378 16-core CPUs running at 2.40 GHz, 256 GB of RAM, and 10 GbE NICs.
The setup 
consists of 5 benchmark nodes and 20 blockchain nodes in addition to up to 20 clients, each sending 
native transfer transactions to a single blockchain node. 
For automation, we use the Diablo~\cite{GGLNV23} benchmark 
where each client is an independent Diablo secondary.

\subsection{Injecting transient packet loss}

  \begin{figure}
  \centering
    \begin{lstlisting}[style=bash, 
                       caption={Commands for injecting transient packet loss.}, 
                       label={lst:packet-loss}]
ethtool -K eth2 tso off gso off gro off sg off
tc qdisc add dev eth2 root handle 1: prio
tc filter add dev eth2 protocol ip parent 1:0 prio 3 u32 match ip dst 10.40.10.1 flowid 1:3
tc qdisc add dev eth2 parent 1:3 handle 30: netem loss 50%
tc qdisc del dev eth2 root
ethtool -K eth2 tso on gso on gro on sg on
    \end{lstlisting}
    \end{figure}

\Cref{lst:packet-loss} indicates how we configured traffic control settings on a network interface to introduce temporary packet losses.
This sequence of commands modifies the network interface \code{eth2} to introduce 50\% packet loss for packets destined for \code{10.40.10.1}, then restores the original configuration.
More precisely, we disabled hardware offloads (line 1), added a \code{prio} qdisc (line 2), filtered traffic to a specific IP  (line 3) into a \code{netem} qdisc applying loss (line 4). We then cleared the rules (line 5) and re-enabled the offloads (line 6).
This method allows for the temporary introduction of packet loss. By dynamically removing the configuration, the effect is made transient without requiring a system restart.

\subsection{Leader detection via logs}\label{ssec:log-monitoring}

We developed a real-time log monitoring system that tracks leaders in blockchain networks. This identifies log patterns to indicate when a node of a leader-based blockchain becomes or ceases to be the leader (e.g., \code{I am now the leader}, \code{ProposerElection}, or protocol-specific block-building messages).
This monitoring system parses node logs 
continuously using \code{inotify}-based file watchers and maintains a stateful model of leader status. Upon detecting a transition into leadership, the monitor triggers a \code{tc} rule to apply packet loss (e.g., \code{loss 75\%}) to the network interface associated with the node and stop this packet loss when the nodes stops being the leader. 
This allows us to isolate node connectivity during their leadership term without modifying the protocol, temporarily degrading their ability to participate in consensus in a targeted way. 

\subsection{Measuring peer-to-peer bandwidth}

We also implemented a fine-grained bandwidth monitoring system to capture the network-level impact of our attacks. Our approach provides pairwise measurements of the traffic exchanged between each node in the network. The system relies on the Linux \code{iptables} firewall infrastructure to perform non-intrusive packet accounting.
For each node under observation, we programmatically install a set of \code{iptables} rules. These rules create custom accounting chains that contain a specific rule for every other peer in the experiment. Each rule is configured to match packets based on their source (for incoming traffic) or destination (for outgoing traffic) IP address.
A monitoring script then periodically queries the byte counters associated with each of these per-peer rules and immediately resets them to zero. This process yields a time series where each data point represents the average transmission (TX) and reception (RX) rate over the preceding interval, allowing us to precisely analyze network behavior.

\subsection{Sending rate}

Limited by the  357\,TPS capacity throughput of Avalanche (induced by the period of 2 second between blocks and their 15M gas limit),
we fixed the sending rate of our experiments to 200\,TPS.
\label{ssec:fee-pb}While this 200\,TPS rate serves as a fair baseline, our initial tests revealed it triggered a load-induced vulnerability in Avalanche dynamic fee mechanism. Under a sustained load, the network gas consumption consistently exceeded its target, causing base fees to escalate until transaction processing halted entirely. To isolate the communication protocol for a fair comparison, we disabled this fee escalation in our genesis configuration for all Avalanche experiments. The specifics of this fee mechanism and our countermeasure are discussed in further detail in \cref{sec:avalanche-target-increase}.

\section{Vulnerability to Targeted Load  Attacks}\label{sec:workload}

In this section, we evaluate the vulnerability of blockchains to targeted load attacks by sending valid, native transfer requests at 200 TPS to only a particular subset of blockchain nodes. 
Similar to DoS attacks against blockchains that were largely studied~\cite{STM18,MJP20,CBR22}, the goal of these attacks is to potentially
increase latency and produce transaction losses.
Interestingly, this attack reveals a bottleneck effect in Aptos delaying its service by one order of magnitude more than other blockchains.
In addition, the attack impacts Avalanche, however, we identified a countermeasure and implemented it by simply disabling Avalanche's new base fee mechanism as explained in \cref{sec:avalanche-target-increase}.

\subsection{Demonstrating latency degradation}

To demonstrate latency degradation, we conducted a failure-free experiment %
with a targeted load attack against the five blockchains.
A single client sends transactions at a sustained rate of 200\,TPS to a single blockchain node. The results revealed a significant performance vulnerability in Aptos. While most blockchains handled the load efficiently, Aptos was an order of magnitude slower. To commit all transactions, Avalanche (base fee disabled) and Redbelly took less than 3 seconds, Algorand took 10 seconds, and Solana took 27 seconds. In stark contrast, Aptos required 4 minutes and 15 seconds. The disparity was also clear in the median (p50) latency, which exceeded 2 minutes and 30 seconds for Aptos while remaining low for the other systems. We note that this attack targets a static node; attacks against the Aptos rotating leader are deferred to \cref{sec:isolation}.

The lack of responsiveness of Aptos is clearly an abnormal behavior. Although Aptos recommends using more resources per node, including 32 cores, 2.8\,GHz and 64\,GB of memory~\cite{apt-reqs}, than what we provisioned in this experiment, we believe that the resources of our experiments are close to those of average commodity computers. In addition, we will show in \cref{ssec:aptos-load-balancing} that our resources are sufficient for Aptos to commit transactions fast (with a median latency of 2.27 seconds) in the absence of this attack.
Because each validator knows the identities of all other validators, this problem uncovers an important vulnerability of Aptos under targeted load: an adversary could potentially 
stop the Aptos service by sending a sustained request rate to a single node.

\subsection{Aptos validator bottleneck}\label{sec:aptos-bottleneck}

\begin{figure}
  \centering
  \includegraphics{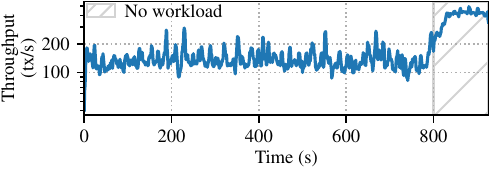}
  \caption{During an 800-second load attack, Aptos congestion prevents it from committing as fast as when the attack stops (after 800 seconds).}
  \label{fig:aptos-none-tps}
\end{figure}

To better understand why Aptos is vulnerable to a targeted load, we measured its throughput during and after the sending of requests.
\Cref{fig:aptos-none-tps} describes the Aptos throughput over time without faults and with a single client sending at 200\,TPS to a single server among 20 servers. 
After 800 seconds, the client stops sending requests but we keep monitoring the throughput during 200 additional seconds as indicated by the striped area.
We observe a sudden jump in throughput after 800 seconds, indicating that Aptos suffered from congestion during the attack sending the 200\,TPS load, seemingly due to an internal bottleneck.

This Aptos bottleneck, causing the vulnerability, is induced by the Quorum Store protocol, discussed in \cref{sec:aptos}. By eliminating the leader bottleneck, the Quorum Store protocol creates a validator bottleneck.

\begin{figure}
\centering
\includegraphics{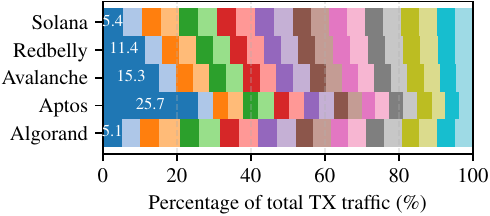}
\caption{Aptos node receiving the transactions generates a quarter of total outgoing traffic during the experiment, up to 5 times higher compared to other blockchains.\label{fig:aptos-bandwidth}}
\end{figure}

To confirm this hypothesis, we analyzed the network traffic distribution across validators. The data reveal a significant concentration of transaction traffic on the node under attack for Aptos. As shown in \cref{fig:aptos-bandwidth}, this single node accounted for 25.7\% of the total outgoing traffic contributed by each node, starkly contrasting with other blockchains. %
This highlights a notable traffic imbalance in Aptos's architecture.

Log inspection across validator nodes
revealed that only the receiving validator node experienced the \code{ProofOfStoreInit} and \code{ProofOfStoreReady} events of the \code{ProofCoordinator} module.
This suggests that only the receiving node initiates and finalizes the proof-of-store process and,
rather than disseminating transactions among validators, the single receiving node must: \textit{(1)}~Process all incoming transactions; \textit{(2)}~Form all batches; \textit{(3)}~Send batches to all other validators; and \textit{(4)}~Collect signatures from $\nicefrac{2n}{3}+1$ validators for each batch.
Since the proof-of-store processing must complete before consensus ordering starts, this node's networking and processing capacity become the primary limiting factors.
As a result, an attacker could load a single Aptos validator node to maximize this bottleneck effect. Distributing the load might seem like a mitigation, but as we show next, it introduces other issues.

\subsection{Balancing the load fails to mitigate attack}\label{ssec:aptos-load-balancing}

\begin{figure}
  \centering
  \includegraphics{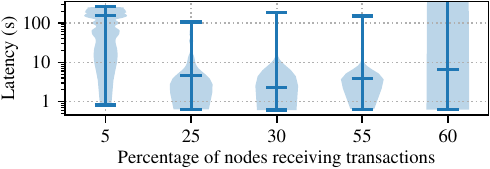}
  \caption{Transaction latency distributions of an Aptos network under a constant uniform workload of 200 TPS for 800 seconds.\label{fig:aptos-sweet-spot}}
\end{figure}

In an attempt to mitigate the targeted attack vulnerability identified above (\cref{sec:aptos-bottleneck}), we varied the number of validator nodes targeted by the 200\,TPS load without injecting faults. 
\Cref{fig:aptos-sweet-spot}
only shows the results when load is balanced on 5\%, 25\%, 30\%, 55\%, and 60\% of the validators as 
Aptos loses transactions when the load is balanced across 60\% to 100\% of the validators. 
This actually means that balancing the load perfectly does not resolve the issue. We observe in \cref{fig:aptos-sweet-spot} that the lowest median latency of 2.27 seconds is achieved when we have 30\% of the validators receiving the transactions.
However, with the 200\,TPS load balanced across 60\% of the validators, Aptos starts losing transactions. With clients sending transactions to all of the nodes, only 50\% of submitted transactions were committed, while with the load balanced across 70\% of the nodes, 90\% of submitted transactions were committed. This indicates that even distributing the targeted load attack can result in a DoS.

We conjecture that the reason that the system is overloaded with too few and too many nodes receiving transactions is due to cryptographic decryption and cryptographic encryption, respectively:
On the one hand, when only one node receives all transactions, it has to verify $\lfloor\nicefrac{2n}{3}\rfloor+1$ signatures for each batch of transactions it creates. The induced CPU consumption slows down the node that then becomes a bottleneck.
On the other hand, when many, say $\Omega(n)$, nodes receive all transactions, then the total number of signatures required becomes quadratic, $\Omega(n^2)$, because different receivers create distinct batches and thus each of these $\Omega(n)$ nodes requires $\lfloor\nicefrac{2n}{3}\rfloor+1$ distinct signatures.

\section{Vulnerability to Transient Failure Attacks}\label{sec:stability}

In this section, we evaluate the vulnerability of blockchains to transient failure attacks. 
These attacks are motivated by the \emph{churn} or participation dynamism either observed in blockchain networks~\cite{IST21} or whose impact is simulated~\cite{MMM20}. We implement such attacks by crashing and recovering nodes, increasing the number of affected nodes in successive experiments. 
Our observations show that Avalanche is vulnerable to these transient failure attacks as it loses transactions 
due to the combination of the failures with its throttling mechanism. Although Aptos seems to also suffer from 
these transient failure attacks, this vulnerability appears to be due to the congestion mentioned in \cref{sec:workload}.
We did not observe any impact on Algorand and Redbelly while transient failures of candidate leaders prevent Solana from stabilizing fast.

\subsection{Avalanche transaction losses}

We anticipate that %
even brief transient failures of a small set of nodes could degrade a blockchain service. To validate this empirically, we produced an attack, initially involving 10\% transient failures, crashing 10\% of the blockchain nodes at 133 seconds while sending transactions exclusively to non-faulty nodes and then recovering the faulty nodes at 266 seconds.
This proportion of transiently failing nodes is relatively low given that blockchains require consensus~\cite{APLPP19} and consensus can be solved in the general setting as long as there are fewer than a third of ``definitive'' failures~\cite{LSP82}.

However, we observe that Avalanche could not tolerate the transient failure attack and started losing transactions. 
In our experiment, Avalanche permanently lost approximately 60\% of the transactions; only 40\% were committed, even long after the faulty nodes had recovered. %
This lasting impact from a brief failure affecting only 10\% of the stake is surprising for two reasons. 
First, 
Avalanche is expected to tolerate the owners of up to 20\% of stake being offline. %
Second, we only injected failures whose nature is less harmful than \emph{Byzantine} (or malicious) failures as they are transient crashes.

\subsection{Avalanche throttling amplification}

\begin{figure}
\centering
\includegraphics{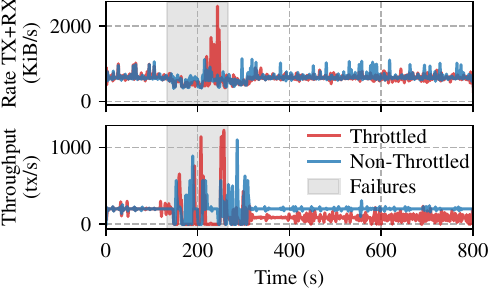}
\caption{Network traffic (TX+RX Rate) and transaction throughput of Avalanche under a transient failure attack. With throttling enabled (red), attempts to sample failed nodes for consensus during the failure window (133-266s) create a large spike in unproductive network traffic. This activity triggers the throttling mechanism, which then suppresses network activity and prevents throughput from recovering post-failure, causing sustained transaction loss. In contrast, the non-throttled system (blue) shows no traffic spike and recovers quickly.\label{fig:stability-bandwidth}}
\end{figure}

To mitigate DoS attacks, Avalanche features a throttling mechanism that rate-limits message processing based on resource consumption. While intended as a defense, this mechanism appears to be the root cause of the instability under transient failures. As shown in \cref{fig:stability-bandwidth}, when we repeated the 10\% transient failure experiment with throttling disabled (blue line), Avalanche successfully recovered and committed 100\% of transactions. With throttling enabled (red line), however, the network failed to recover, leading to many transaction losses.

This amplification effect stems from consensus sampling failures, not from a backlog of transactions. During the failure period, Avalanche's dynamic proposer selection protocol continues to sample from the entire validator set, including the 10\% of nodes that are offline. As healthy nodes attempt to communicate with these unresponsive, downed proposers, they generate a large volume of unproductive network traffic, seen as the sharp red spike in the top panel of \cref{fig:stability-bandwidth}.

The throttling mechanism misinterprets this spike of failed consensus communication as a malicious overload and begins to suppress message queues. Consequently, even after the failed nodes recover at 266s, the throttling defense remains active, rate-limiting the legitimate communication needed to resume consensus. This prevents the system from recovering its throughput, as shown in the bottom panel of \cref{fig:stability-bandwidth}, and results in the permanent loss of transactions submitted during and after the attack. Our finding clarifies a previous observation that throttling could create throughput instability~\cite{GGL24}; however, that result was obtained on the default configuration where, as we explain in \cref{sec:discussion}, the dynamic fee mechanism can also cause halts. Our work demonstrates that throttling is a distinct vulnerability, causing transaction loss even after the fee issue is corrected.

\subsection{Aptos transaction losses}

Aptos is also vulnerable to transient failure attacks %
but requires a much higher percentage of failures to be impacted. We ran a more severe attack against Aptos, crashing 35\% of its blockchain nodes at 133 seconds and recovering them at 266 seconds. As transactions were taking longer to be committed, probably due to the validator bottleneck presented in \cref{sec:aptos-bottleneck}, we had to increase the duration of our experiments.
Even under this harsher 35\% failure scenario, Aptos proved more resilient than Avalanche, eventually committing 89\% of its transactions. %
However, the performance was significantly degraded, requiring over 600 seconds to process the workload, and still resulted in a final loss of 11\% of submitted transactions. Recall that this is the result of injecting transient failures when all transactions are sent to non-faulty nodes.

\subsection{Avalanche impact of multiple failures}
\label{ssec:avalanche-transient-faults}

\begin{figure}
\centering
\includegraphics{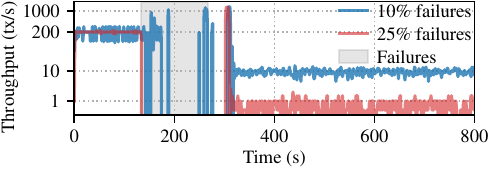}
\caption{Avalanche throughput drop as the number of transient failures increases.\label{fig:avalanche-transient-faults-evolution}}
\end{figure}

Interestingly, Avalanche throughput decreases as more nodes are targeted by the transient failure attack. This progressive degradation is less pronounced in Algorand, Redbelly, and Solana.
\Cref{fig:avalanche-transient-faults-evolution} 
depicts the throughput over time, averaged over a 2-second sliding window, of an Avalanche network receiving 200\,TPS for 800 seconds. 
The blue (resp. red) curve depicts an experiment where 10\% (resp. 25\%) of nodes fail at 133 seconds and recover at 266 seconds. 
Before the failures, throughput remains relatively stable.
At 133 seconds, the red curve experiences a sharper drop than the blue curve.
After the failed nodes recover, the throughput remains unexpectedly low:
with 2 failures, only 36\% of transactions are committed, while with 5 failures, only 22\% of transactions are committed.

The reason for this vulnerability is a cascading effect when the mempool is full: rejecting transactions creates the later rejection of subsequent transactions with higher nonces. 
About 26,600 transactions could be submitted during the attack (200\,TPS $\times$ 133\,s), while Avalanche transaction pool can only store 6,144 transactions.
As transactions get rejected by lack of space, newer transactions with higher nonces also get rejected with an \code{ErrUnderpriced} error, despite being equally priced, due to \code{pool.priced.Underpriced(tx)=true}. 
Additionally, recovering nodes have an empty mempool and must receive transactions from other nodes. The more failed nodes, the more transaction losses, exacerbating the impact on throughput.
When the attack targets 25\% of the network, Avalanche performance degrades further, which is explained by its $\nicefrac{n}{5}$ failure threshold depicted in \cref{table:comparison}.

\section{Vulnerability to Packet Loss Attacks}\label{sec:messageloss}

\definecolor{highlightcolor}{gray}{0.9}

\begin{table*}[ht!]
\centering
\caption{Comparison of average peer-to-peer bandwidth of a peer from $\lfloor f+1 \rfloor$ group before, during, and after a 50\% packet loss attack}
\label{tab:bw-comparison-full}
\setlength{\tabcolsep}{1pt} 
\begin{tabularx}{\textwidth}{l *{6}{>{\raggedleft\arraybackslash}X}}
\toprule
 & \multicolumn{3}{c}{\textbf{Avg. Peer TX Rate (KiB/s)}} & \multicolumn{3}{c}{\textbf{Avg. Peer RX Rate (KiB/s)}} \\
\cmidrule(lr){2-4} \cmidrule(lr){5-7}
\textbf{Blockchain} & \textbf{Before} & \textbf{During (\% $\Delta$)} & \textbf{After (\% $\Delta$)} & \textbf{Before} & \textbf{During (\% $\Delta$)} & \textbf{After (\% $\Delta$)} \\
\midrule
Algorand & 1,480 & 67 (-95.5\%) & 1,373 (-7.3\%) & 1,550 & 59 (-96.2\%) & 1,483 (-4.4\%) \\
Aptos & 112 & 18 (-83.8\%) & 68 (-39.0\%) & 162 & 16 (-90.1\%) & 247 (+51.9\%) \\
Avalanche & 152 & 24 (-84.2\%) & 134 (-12.1\%) & 172 & 20 (-88.5\%) & 155 (-9.5\%) \\
Redbelly & 547 & 24 (-95.6\%) & 571 (+4.3\%) & 534 & 23 (-95.7\%) & 617 (+15.5\%) \\
\rowcolor{highlightcolor} Solana & 840 & 840 (-0.0\%) & 824 (-1.8\%) & 872 & 595 (-31.8\%) & 865 (-0.7\%) \\
\bottomrule
\end{tabularx}
\end{table*}

In this section, we evaluate the vulnerability of blockchains to packet loss attacks, where an adversary intentionally drops network messages between nodes or network segments. 
Similarly to the balance attacks~\cite{NG17,NTT21,NEJ24} that aim at partitioning the network into balanced subnetworks, such attacks disrupt communication between two subnetworks. We execute these attacks by injecting varying percentages of packet loss. While achieving high packet loss network-wide is challenging, these tests reveal inherent protocol sensitivities. We find that the choice of transport protocol (TCP vs. \quic) significantly affects tolerance to packet loss, a conclusion strongly supported by the peer-to-peer bandwidth measurements shown in \cref{tab:bw-comparison-full}. The data reveals a dramatic impact on TCP-based protocols: during the attack, average peer bandwidth for chains like Algorand and Redbelly collapses by over 95\%. In stark contrast, Solana, which uses \quic, maintains its full transmission (TX) rate throughout the attack and sees only a partial reduction in receive (RX) bandwidth. This network-level resilience prevents the cascading failures, such as full broadcast queues, that affect TCP-based systems like Algorand.

\subsection{Attacking with packet losses}

To evaluate the vulnerability to packet loss attacks, we drop between 25\% and 75\% of the packets exchanged between a group of 
$\lfloor f+1 \rfloor$ nodes, where $f$ is the fault tolerance ratio of each blockchain reported in \cref{table:comparison},
and the rest of the network after 133 seconds. Finally, we stop losing packets after 266 seconds.
To avoid the Aptos congestion observed in \cref{ssec:aptos-load-balancing}, we select 25\% of the blockchain nodes for all experiments of this section.

\subsection{Vulnerability of TCP-based blockchains}

\begin{figure}
\centering
\includegraphics{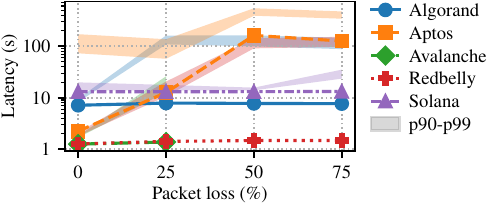}
\caption{Transaction latency percentiles of the 5 blockchains under test when stressed with constant 200\,TPS workload for 800 seconds. %
Lines represent median latency.\label{fig:dissemination-comparison}}
\end{figure}

\Cref{fig:dissemination-comparison} illustrates the effect of increasing packet losses on the latency of the 5 blockchains. The x-axis represents the percentage of packet losses introduced into the system, while the y-axis, plotted on a logarithmic scale, denotes latency in seconds. Each blockchain p50 (i.e., the median latency) is shown as a solid line, while the shaded regions capture the variability between p90 (i.e., the $90^{\ms{th}}$ percentile) and p99 (i.e., the $99^{\ms{th}}$ percentile) latencies.
With fewer (25\%) packet losses, 
the vulnerability is more pronounced for Aptos and Algorand, which experience a significant rise in p90 and p99 latencies.
With more (50\% and 75\%) packet losses, 
however, a more noticeable divergence appears. 
The most extreme vulnerability is seen in Avalanche, for which p50, p90, and p99 values are missing at 50\% and 75\% loss.
Similarly, the sharp latency increase for Aptos and Redbelly at these loss levels suggests severe delays in finalization due to stalled consensus under adverse network conditions.

Algorand, like other TCP-based blockchains, is even more impacted during the 50\% and 75\% packet loss attacks, which means transactions are not committed during the attack. This results in an apparent gap where no transactions finalize, followed by a significant latency increase as backlogged transactions begin committing after the losses stop. Notably, Algorand also exhibits a slower recovery compared to other blockchains, taking longer to regain its previous throughput levels once packet loss ceases. This delayed recovery further contributes to the observed increase in latency variability, as transactions submitted during the loss period accumulate additional delays before finalization.

\subsection{Algorand transaction loss}

Algorand is particularly impacted by even 25\% packet loss. The reason is threefold. First, Algorand stores transactions in a mempool via the \code{Remember} function and stores transactions to send in a broadcast queue: the packet loss leads transactions to remain unprocessed in the broadcast queue leading to logged messages of the type \code{HTTP 400: {"message":"broadcast queue full"}}. When the broadcast queue is full, new messages that cannot be enqueued are dropped.  Second, once connectivity is restored, we observed that Algorand needed an additional 99 seconds to resume transaction processing, which confirms previous observations~\cite{GGL24}. Third, Algorand could potentially retrieve the missed transactions, however, its \code{TxSyncer} synchronization protocol builds upon a Bloom filter whose false positives may result in transactions not being requested, leading to previously mentioned ``stuck'' transactions~\cite{algorand-sync-issue}.
The combination of these three phenomena prevents some transactions from being eventually committed.

\subsection{Multi-layered resilience in Solana}\label{ssec:hierarchical}

In contrast with TCP-based protocols, Solana demonstrates significantly higher resistance to packet loss (\Cref{fig:dissemination-comparison}). This resilience stems from a multi-layered strategy. First, it uses \quic instead of TCP, a choice that provides reliability and flow control while avoiding the head-of-line blocking that can stall data streams during packet loss.

Second, its Turbine block propagation protocol attacks the problem directly. Turbine disseminates blocks through a hierarchical tree, and critically, uses erasure coding to split blocks into fragments (``shreds'') with added redundancy. This allows validators to reconstruct full blocks even if a significant fraction of fragments are lost in transit. Finally, a stake-weighted quality of service (QoS) mechanism allocates network bandwidth proportional to validator stake, preventing congestion caused by spam and ensuring fair resource access. Together, these mechanisms make Solana inherently more robust against network disruption.

\section{Vulnerability to Stopping Attacks}\label{sec:liveness} %

In this section, we investigate the vulnerability of blockchains to ``stopping attacks''. Like in majority attacks where a validator can impose its block~\cite{Ros12}, grow a large coalition~\cite{ES18} or own most computational power~\cite{AFO21}, these attacks rely on an adversary controlling a majority of nodes, but to stop the network. This inability to resume normal operations even after the faulty nodes rejoin the network distinguishes such attacks from the transient failure attacks of \cref{sec:stability}.
We observe that large-scale attacks cause Solana to halt indefinitely even after all nodes recover, while Avalanche experiences a near-halt state, unable to recover at least the transactions issued during the transient failures.

\subsection{Avalanche near-stopping}\label{sec:solana-avalanche-stop}

\begin{figure*}
\centering
\includegraphics{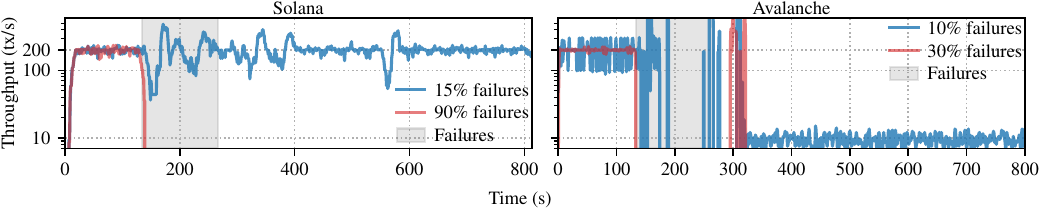}
\caption{Differences of Solana throughput variations with 15\% and 90\% transient failures and of  Avalanche throughput variations with 10\% and 30\% transient failures. %
Solana crashes due to transient failures while Avalanche cannot recover its initial throughput after fewer transient failures.
\label{fig:solana-transient-faults1}}
\end{figure*}

We know from \cref{ssec:avalanche-transient-faults} that the combination of transaction nonce management and throttling prevents Avalanche from committing some transactions.
\Cref{fig:solana-transient-faults1} contrasts Avalanche throughput over time, computed over a 2-second sliding window, under a  transient failure attack of 10\% versus of 30\% of the network from the $133^{\ms{rd}}$ to the $266^{\ms{th}}$ second. With 10\% of transient failures, throughput is significantly reduced but does not completely halt. However, with 30\% transient failures, throughput drops sharply and does not recover even after the failed nodes restart. This is due to the compounding effects of lost transactions, nonce gaps, and throttling, which together prevent transaction execution.
Importantly, this issue is not fundamental to the Avalanche consensus mechanism but rather an artifact of its transaction pool configuration. We found that increasing the transaction pool size and disabling throttling during recovery mitigates this specific near-stopping vulnerability and allows the system to regain its previous throughput, demonstrating that these factors are responsible for the observed halt.

\subsection{Solana vulnerability to stopping attacks}\label{ssec:solana-no-commit}

\begin{figure}
\centering
\includegraphics{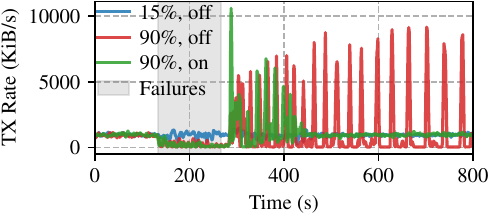}
\caption{Outgoing bandwidth of the node receiving transactions from the client demonstrates transaction propagation patterns. Without the flag, uncommitted transactions continuously re-propagate through the mempool-less forwarding mechanism, creating sustained bandwidth spikes. With the flag, bandwidth normalizes after recovery as the transaction backlog is processed. The 15\% failure case shows normal operation.
\label{fig:solana-transient-faults1-bandwidth}}
\end{figure}

Solana is vulnerable to a stopping attack where a simultaneous transient failure of a large fraction of its nodes (e.g., 90\%) causes a complete and unrecoverable halt in transaction processing, as shown in \cref{fig:solana-transient-faults1}. This stall manifests not only in zero throughput but also in characteristic network behavior: \cref{fig:solana-transient-faults1-bandwidth} shows how uncommitted transactions continuously re-circulate through Solana transaction forwarding component, creating persistent bandwidth spikes that subside only after recovery.

The root cause is a liveness stall related to its vote lockout mechanism. After a mass restart, our log analysis revealed that newly elected leaders refuse to produce blocks because they are programmed to wait for a ``rooted vote'' from a supermajority. However, since the supermajority is offline and the few surviving nodes cannot produce rooted blocks on their own, the network enters a deadlock. Restarted nodes wait for a condition that cannot be met, leading to an indefinite stall.

We confirmed that this vulnerability is not fundamental to the consensus protocol, but stems from a default configuration. By enabling the \code{-{}-no-wait-for-vote-to-start-leader} flag, which allows leaders to begin block production without observing a recent rooted vote, the network successfully recovered after the attack. This shows that the strict dependency on rooted votes creates a single point of failure during mass-recovery events.

\subsection{Resistance against stopping attacks}

Additionally, we ran the same stopping attacks on Algorand, Aptos and Redbelly and with different proportions of the network experiencing transient failure attacks. We made two observations:
\begin{enumerate}
\item Algorand, Aptos and Redbelly proved resilient to stopping attacks under our test conditions: even after a 95\% stopping attack, these blockchains were able to recover liveness and commit newly issued transactions after the nodes restarted. 
\item Avalanche would almost halt after at least 25\% transient failures while Solana would halt after at least 85\% transient failures. In our experiments Solana would tolerate proportions of transient failures from 15\%, 20\%, ..., 80\%. It would, however, halt in certain executions with 85\% transient failures. And the same with 90\% and 95\% transient failures. 
\end{enumerate}

\section{Vulnerability to Leader Isolation Attacks}\label{sec:isolation}

\begin{figure}
\centering
\includegraphics{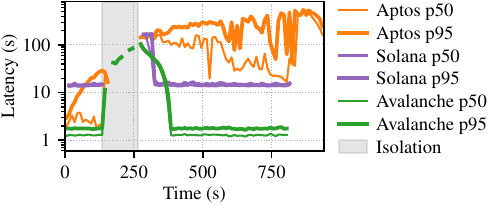}
\caption{Latency percentiles (p50 thin solid line, p95 thick solid line) calculated over 10-second bins based on transaction commit times for Aptos, Solana, and Avalanche under leader isolation. Gaps in lines correspond to bins with zero committed transactions (zero throughput). The shaded region indicates the leader isolation period (133s to 266s).\label{fig:isolation}}
\end{figure}

We introduce a new set of attacks, called Leader isolation attacks, where
one individual node at a time is isolated from the rest of the network.
It resembles the isolation of the eclipse attacks conducted against Bitcoin~\cite{HKZG15},
Ethereum~\cite{WG16} and Monero~\cite{SPL25}
but instead it targets a potential leader of the underlying consensus protocol. 
The goal of these attacks is 
to determine whether delaying the communication of a single node executing the consensus protocol can disrupt the blockchain progress without the need for a large network partition.

We observe that this form of targeted isolation limits progress of Aptos and Solana and degrades the performance of Avalanche.
We omit the results on Algorand and Redbelly because these blockchains exploit randomized and leaderless consensus protocols, respectively. These consensus protocols do not rely on the timely communication of a single node or ``leader'' to coordinate progress but instead replicate this task at multiple nodes to naturally cope with a single point of failure.

\subsection{Performing leader isolation attack}\label{ssec:isolation-attack-descr}

For each blockchain, we evaluate the impact of a leader isolation attack under a uniform experimental configuration. Transactions are submitted to the network at a sustained rate of 200\,TPS over a total duration of 800 seconds.

To simulate the attack, we identify the current consensus leader using real-time log monitoring (\cref{ssec:log-monitoring}). Once a node is detected to be in the leader role, we apply a targeted network impairment by injecting packet loss using the Linux \code{tc} subsystem. Specifically, we configure a \code{netem} rule to drop 75\% of both incoming and outgoing packets at the leader node during its term.

The attack window begins at 133 seconds into the experiment and ends at 266 seconds. Outside this window, network conditions are left unmodified. The selective nature of the attack ensures that the rest of the network remains unaffected, allowing us to assess the impact of impaired leader connectivity on consensus progress and transaction processing.

\subsection{Aptos stops and struggles to recover}

\begin{figure}
\centering
\includegraphics{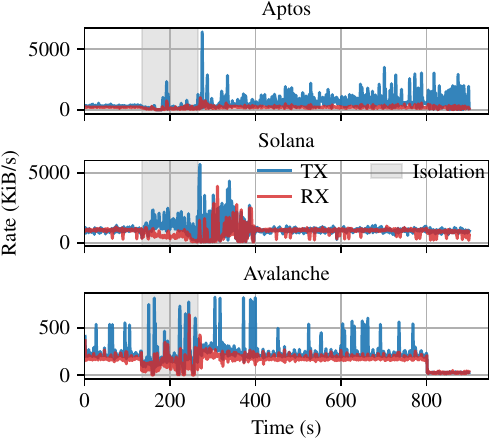}
\caption{The TX and RX rates of a single node receiving transactions from the client. Aptos experiences a massive, sustained TX spike post-attack, correlating with the high recovery latency in \cref{fig:isolation}. Solana shows a brief burst of activity before quickly returning to normal levels, indicating efficient backlog processing. Avalanche's bandwidth is only moderately impacted during the attack and recovers smoothly, consistent with its non-stopping behavior.\label{fig:isolation-bandwidth}}
\end{figure}

As shown in \cref{fig:isolation}, Aptos stops as long as its leader is isolated.
More precisely, the median and p95 (i.e., $95^{\ms{th}}$ percentile) latencies of Aptos are relatively stable before the attack.
However, the number of committed transactions drops to zero soon after the attack starts and then remains null until the attack stops. This is illustrated by the interruption of the Aptos p50 and p95 latency curves few seconds after second 133 and until second 266. The network-level impact of this halt and subsequent recovery struggle is starkly illustrated in \cref{fig:isolation-bandwidth}. While the leader is isolated, network activity is minimal. However, once the attack ends at 266s, the node's transmission (TX) rate spikes dramatically and remains elevated for the remainder of the experiment.
Our investigation of the output logs confirms that Aptos stops. First, we found an $\lit{INFO}$ message stating \code{Quorum store is back pressured with}, suggesting that the transaction pool receives transactions faster than it can execute them. Second, messages \code{Executing block, transaction count} are logged prior to and after the attack but stop being logged during the attack, demonstrating the corresponding lack of progress. 
To conclude, these experiments confirm that isolating the leader is sufficient to stop the Aptos service.

Note that Aptos reputation-based leader election cannot help as the attack can predict and target each subsequent leader.
In particular, isolating the current leader creates a vicious circle: 
Because an isolated leader cannot propose correctly, 
its  \code{failed\_proposals}
count increases, which contributes to its failure rate. When its failure rate reaches 10\%, the chances of this node being newly elected as a leader of an upcoming round are divided by 1000.
As this election is deterministic, the continuous attack on the newly elected leader counteracts the reputation-based election by lowering the reputation of each  potential leader one by one, hence contributing to stopping the whole Aptos network.

\subsection{Solana stops but recovers quickly}

Similarly to Aptos, \cref{fig:isolation} shows that Solana stops while its leader is isolated. This is illustrated by the p50 and p95 latency curves disappearing completely during the attack (in the shaded area).
The attack targets each leader in turn as specified by the Solana's leader schedule (cf. \cref{ssec:solana}) and prevents the Turbine protocol (cf. \cref{ssec:hierarchical}) from disseminating blocks and receiving Tower BFT votes.
We observed this problem in the logs with the disappearance of \code{new root} messages, indicating the lack of votes from a supermajority, which are needed to finalize blocks and advance the blockchain's root.
In addition, the presence of \code{Waiting to switch vote to...} logged messages suggests that validators observe heavier forks without being able to commit to them or make progress.

When the leader isolation attack stops, Solana latency curves reappear. The next designated leader in the schedule can successfully create its blocks and the Tower BFT consensus protocol resumes. 
The latency, and in particular the p95, spike, due to the processing of the old backlog of transactions.
Our observation of the transaction timestamps indicates that the transactions submitted during the attack experience a delay
of about the attack duration plus the processing time, $\simeq 150$ seconds, which contributes to this peak.
After clearing the backlog, Solana experiences a rapid recovery observable by p95 getting close to p50. This quick recovery is mirrored in Solana's network bandwidth profile, shown in \cref{fig:isolation-bandwidth}. Following the isolation period, both TX and RX rates spike as the new leader disseminates blocks and clears the backlog. Crucially, this burst of activity is short-lived, and network traffic quickly returns to pre-attack levels.
To conclude, similarly to Aptos, Solana stops as soon as its leader is isolated. However, in contrast with Aptos and the load induced by Quorum Store (cf. \cref{sec:aptos-bottleneck}), Solana recovers quickly after the isolation attack stops by having latencies getting close to normal conditions.

\subsection{Avalanche progresses with increased latency}

In contrast with Aptos and Solana, Avalanche does not completely stop during the isolation attack. In fact, \cref{fig:isolation} shows that the latency curves, although interrupted, appear in the shaded area of the attack period.
This is due to Avalanche's ``soft'' proposer mechanism (cf. \cref{ssec:avalanche}):
As the designated proposers fail to produce a block within their collective windows due to the attack, the protocol allows any active validator to propose
after the windows.
Since our isolation attack against Avalanche consists of impairing a validator as soon as it is detected as a proposer and during 5 seconds (which corresponds to the Avalanche's default \code{WindowDuration}), there is a chance that another validator successfully propagates a block when not impaired.
This is why performance is impacted but progress is not stopped. The network bandwidth data in \cref{fig:isolation-bandwidth} further supports this observation of continued, albeit degraded, progress. Unlike Aptos and Solana, Avalanche does not halt communication; instead, its TX and RX rates show only a moderate decrease during the isolation period. After the attack ends, there is no significant traffic spike. The network smoothly returns to its normal operational state, indicating that Avalanche's protocol successfully avoids building a large transaction backlog by continuing to process work, even when designated proposers are offline.

When the attack is active, any node identified as attempting to propose -- whether a designated proposer acting within its window or \emph{any} validator attempting to build after the windows expire -- experiences significant network impairment as described in \cref{ssec:isolation-attack-descr}. This impairment hinders the affected node's ability to successfully propagate its block and gather necessary confirmations. Logs from the attack period confirm communication difficulties, showing messages indicating that requests were \code{throttled} or \code{timed out}, likely due to the induced packet loss disrupting network interactions.
We observe that both the p50 and p95 latencies increase during the attack
and remain high after the attack. This suggests lingering network effects or backlogs caused by the period of inefficient and impaired block production.

\section{Discussion and Countermeasures}\label{sec:discussion}

In this section, we discuss interesting aspects and countermeasures of the gas fee management of Avalanche and the warmup of Solana that directly impacts the services when failures occur.

\subsection{Gas fee impact on Avalanche performance}\label{sec:avalanche-target-increase}

Avalanche C-Chain gas fees are dynamic~\cite{ava-fees}. Initial dynamic fees (Apricot Phase 3~\cite{ava-apr-3}) evolved with Snowman++ and block-based fees (Phase 4~\cite{ava-apr-4}), eventually leading to an increased gas target of 15M gas every 10 seconds (Phase 5~\cite{ava-apr-5}) to improve scalability~\cite{ava-block-explorer,ava-galaxy}. While beneficial on mainnet, this dynamic fee mechanism caused issues in our sustained 200\,TPS benchmark: target gas consumption consistently exceeded 15M, leading to constantly increasing base fees until transactions could no longer be submitted~\cite{ava-rpc-fee-cap}. As explained in \cref{ssec:fee-pb}, we identified a simple countermeasure to this problem, which consists of bypassing this escalation fee. We also validated this countermeasure experimentally in \cref{sec:messageloss,sec:liveness,sec:isolation}.
Consequently, Avalanche's performance under attacks should be interpreted in the context of this specific genesis configuration, which was necessary to probe deeper protocol behaviors beyond the initial fee-related halt.

\subsection{Avoiding warmup crashes in Solana}\label{ssec:solana-warmup-problem}

We noticed that the problem of Solana's global outage that we discovered in \cref{ssec:solana-no-commit} impacts the warmup phase of the latest version of the available source code. However, we also noted that this problem is absent from the version in production.

More specifically, the default local network deployment scripts in Solana enable warmup, which makes the network start with small epoch sizes, and gradually increase to the normal size. This caused the network to crash when failures were introduced because of the hash calculation explained in detail in~\cite{GGL24}.

Although the halting problem we discovered in 
\cref{ssec:solana-no-commit} remains and prevents Solana from being live under transient failures, a simple countermeasure consists of disabling the warmup. In particular, we validated empirically that Solana's global outage of \cref{ssec:solana-no-commit} no longer happens with our countermeasure.

\section{Related Work}\label{sec:relatedwork}

Our packet loss attacks are similar to different partitioning attacks specifically tailored for Bitcoin~\cite{HKZG15}, Ethereum~\cite{NEJ24} 
and ZLB~\cite{RPG24}, however, ours control precisely the proportion of packet losses.
Our leader isolation attacks build on the known observation that leader-based consensus protocols are vulnerable to the failure of a single node~\cite{ABDPGGVZ23},  
a centralization that is known to put blockchains at risk~\cite{vSRGG25}. 
This leader isolation can be viewed as a variant of the eclipse attack against Bitcoin~\cite{HKZG15} but adapted for leader-based blockchains.
Our targeted load attack shares similarities with some DoS attacks~\cite{HWYKS23}, however, the former overloads Aptos while the latter partitions Ethereum. Other DoS attacks target transaction execution logic and fee markets~\cite{YQZ24}, RPC services~\cite{LCL21}, Ethereum's txpool~\cite{LWT21}, amplification attacks~\cite{TZQ25}, and specific protocol vulnerabilities like TON's ADNL~\cite{FK25}. Our study complements these by focusing on broader network-level attacks to compare foundational communication resilience across multiple blockchains.

Network topology plays a crucial role in blockchain resilience. Studies~\cite{LTC21,ZZZ24,DPSF25,DPBF25} have provided significant insights into network structures of various blockchains. While our research does not involve topology discovery, these works highlight network characteristics that can influence the impact of the attacks we investigate.

Experimenting vulnerabilities directly on isolated blockchain networks is becoming more and more common. 
The Blockchain Anomaly~\cite{NG16} and the Balance Attack~\cite{NG17} were demonstrated on networks of machines running the original Ethereum protocol with proof-of-work. 
The {\sc Stabl} framework~\cite{GGL24} injects failures on blockchain networks to assess their ``sensitivity'', however, the sensitivity metric is insufficient for our comparisons:
(i)~it increases identically whether failures benefit or deteriorate performance compared to a baseline and (ii)~becomes infinite as soon as the blockchain loses one transaction.

The reliability of replicated state machines
was previously assessed by injecting crashes~\cite{VG20} or Byzantine failures~\cite{singh_bft_2008,lee_turret_2014,bano_twins_2022,AAA24} without considering the overall blockchain protocols.
The reliability of blockchain protocols was previously assessed by injecting Byzantine failures~\cite{CNG21,RPG24}, injecting system call failures~\cite{Zhang23}, using fuzzing~\cite{YTB21,ma_loki_2023,CMZ23,winter_randomized_2023,MaPhoenix23,ZBJ25} or injecting network delays~\cite{GJH23,RPG24}.
They cannot compare the fault tolerance of different blockchains on the same ground.

Most comparisons of blockchain protocols typically focus on evaluating performance in the absence of attacks~\cite{dinh_blockbench_2017,nasrulin_gromit_2022,GGLNV23,GJH23,ma_gfbe_2024,noauthor_hyperledger_2024}. They do not measure the vulnerabilities of blockchain protocols in adversarial conditions.
Recent research results identified limitations in Avalanche~\cite{ACS24} and Solana~\cite{SKW24}.
The observations are different from ours.
In~\cite{ACS24} the authors provide a probabilistic analysis 
of Avalanche 
but do not explicitly indicate how some attack or its 
gas fees can impact its progress.
In~\cite{SKW24} the authors demonstrate how a single malicious leader in Solana can halt the protocol but do not measure the impact of message losses or transient failures.
In~\cite{GGL24} the authors show how to crash Solana, however, this crash results directly from the publicly available scripts and cannot be generalized to the in-production version as we explained in \cref{ssec:solana-warmup-problem}.

\section{Conclusion}\label{sec:conclusion}

In this paper, we presented the first comparative vulnerability analysis of blockchain communication protocols. We studied five modern blockchains, Algorand, Aptos, Avalanche, Redbelly, Solana, under targeted load, transient failure, packet loss, stopping, and leader isolation attacks, which led to the overview of~\vref{table:comparison}. 
More precisely, our findings reveal that Aptos latency increases severely under targeted load due to its validator bottleneck; Avalanche and Aptos lose transactions under transient failures;
TCP-based protocols are more vulnerable to packet loss attacks than \quic-based protocols combined with erasure coding; 
Solana can crash after an excessive amount of transient failures while Avalanche also showed near-stopping vulnerabilities; leader isolation attacks stopped Aptos and Solana and degraded Avalanche, highlighting risks in leader-based designs. 
We also provided countermeasures to avoid the identified outages: To cope with Avalanche's outage, one can cap the base fee increase and to cope with Solana's outage one can disable its warmup.
We plan to open source our framework and test other blockchains.

\section*{Acknowledgment}

This
research is supported under Australian Research Council Discovery Project funding scheme (project number 250101739) entitled ``Fair Ordering of Decentralised Access to Resources''.

\bibliographystyle{plainurl}
\bibliography{references}

@inproceedings{GGL24,
  author    = {Gramoli, Vincent and Guerraoui, Rachid and Lebedev, Andrei and Voron, Gauthier},
  title     = {Stabl: The Sensitivity of Blockchains to Failures},
  year      = {2025},
  booktitle = {Proc. Middleware},
  pages     = {202--214}
}

@misc{Sol25,
  title  = {{S}olana network outages as \${TRUMP}, \${MELANIA} and other {S}olana-based memecoins surge},
  author = {Cryptodaily},
  url    = {https://cryptodaily.co.uk/2025/01/solana-network-outages-as-trump-melania-and-other-solana-based-memecoins-surge},
  year   = {2025},
  note   = {Accessed: Jan. 20, 2025}
}

@inproceedings{GHM17,
  author    = {Gilad, Yossi and Hemo, Rotem and Micali, Silvio and Vlachos, Georgios and Zeldovich, Nickolai},
  title     = {{A}lgorand: Scaling {B}yzantine Agreements for Cryptocurrencies},
  booktitle = {Proc. ACM SOSP},
  year      = {2017},
  pages     = {51--68}
}

@misc{Yak21,
  author = {Anatoly Yakovenko},
  year   = {2021},
  title  = {{S}olana: A new architecture for a high performance blockchain v0.8.13},
  url    = {https://solana.com/solana-whitepaper.pdf},
  note   = {Accessed: Aug. 31, 2024}
}

@inproceedings{CNG21,
  author    = {Tyler Crain and Christopher Natoli and Vincent Gramoli},
  title     = {{R}ed {B}elly: a Secure, Fair and Scalable Open Blockchain},
  booktitle = {Proc. IEEE S\&P},
  year      = {2021}
}

@misc{noauthor_aptos_2022,
  author = {Aptos},
  title  = {The {Aptos} Blockchain: Safe, Scalable, and Upgradeable {Web3} Infrastructure},
  url    = {https://aptosfoundation.org/whitepaper/aptos-whitepaper\_en.pdf},
  note   = {Accessed: Aug. 31, 2024},
  year   = {2022}
}

@techreport{rocket_scalable_2020,
  title       = {Scalable and Probabilistic Leaderless {BFT} Consensus through Metastability},
  institution = {arXiv},
  author      = {Rocket, Team and Yin, Maofan and Sekniqi, Kevin and van Renesse, Robbert and Sirer, Emin Gün},
  year        = {2020}
}

@inproceedings{GGLNV23,
  author    = {Vincent Gramoli and Rachid Guerraoui and Andrei Lebedev and Chris Natoli and Gauthier Voron},
  title     = {{Diablo}: A Benchmark Suite for Blockchains},
  booktitle = {Proc. EuroSys},
  pages     = {540--556},
  year      = {2023},
}

@inproceedings{SKW24,
  author    = {Jakub Sliwinski and Quentin Kniep and Roger Wattenhofer and Fabian Schaich},
  title     = {Halting the {S}olana blockchain with epsilon stake},
  booktitle = {Proc. ICDCN},
  pages     = {45--54},
  year      = {2024}
}

@inproceedings{ACS24,
  author    = {Amores-Sesar, Ignacio and Cachin, Christian and Schneider, Philipp},
  title     = {An Analysis of {A}valanche Consensus},
  year      = {2024},
  booktitle = {Proc. SIROCCO},
  pages     = {27--44}
}

@inproceedings{GJH23,
  author    = {Geyer, Frank Christian and Jacobsen, Hans-Arno and Mayer, Ruben and Mandl, Peter},
  title     = {An End-to-End Performance Comparison of Seven Permissioned Blockchain Systems},
  year      = {2023},
  booktitle = {Proc. Middleware},
  pages     = {71--84}
}

@inproceedings{nasrulin_gromit_2022,
  title     = {Gromit: Benchmarking the Performance and Scalability of Blockchain Systems},
  booktitle = {Proc. IEEE DAPPS},
  author    = {Nasrulin, Bulat and De Vos, Martijn and Ishmaev, Georgy and Pouwelse, Johan},
  year      = {2022},
  pages     = {56--63}
}

@article{ma_gfbe_2024,
  title   = {{GFBE}: A Generalized and Fine-Grained Blockchain Evaluation Framework},
  volume  = {73},
  number  = {3},
  journal = {IEEE Trans. Comput.},
  author  = {Ma, Liyuan and Liu, Xiulong and Li, Yuhan and Zhang, Chenyu and Shi, Gaowei and Li, Keqiu},
  year    = {2024},
  pages   = {942--955}
}

@misc{noauthor_hyperledger_2024,
  author = {Dave Kelsey},
  title  = {Hyperledger {Caliper}},
  url    = {https://hyperledger.github.io/caliper/},
  note   = {Accessed: Aug. 31, 2024},
  year   = {2024}
}

@inproceedings{dinh_blockbench_2017,
  title     = {{BLOCKBENCH}: A Framework for Analyzing Private Blockchains},
  booktitle = {Proc. ACM SIGMOD},
  author    = {Dinh, Tien Tuan Anh and Wang, Ji and Chen, Gang and Liu, Rui and Ooi, Beng Chin and Tan, Kian-Lee},
  year      = {2017},
  pages     = {1085--1100}
}

@inproceedings{CGT19,
  author    = {Conti, Mauro and Gangwal, Ankit and Todero, Michele},
  title     = {Blockchain Trilemma Solver {A}lgorand has Dilemma over Undecidable Messages},
  year      = {2019},
  booktitle = {Proc. ARES}
}

@misc{IETF21,
  title  = {{RFC} 9000 -- {QUIC}: A {UDP}-Based Multiplexed and Secure Transport},
  author = {IETF},
  year   = {2021}
}

@inproceedings{castro_practical_1999,
  title     = {Practical {Byzantine} fault tolerance},
  booktitle = {Proc. OSDI},
  author    = {Castro, Miguel and Liskov, Barbara},
  year      = {1999},
  pages     = {173--186}
}

@inproceedings{BGL22,
  title     = {Holistic Verification of Blockchain Consensus},
  author    = {Nathalie Bertrand and Vincent Gramoli and Marijana Lazić and Igor Konnov and Pierre Tholoniat and Josef Widder},
  booktitle = {Proc. DISC},
  year      = {2022}
}

@inproceedings{CGLR18,
  author    = {Tyler Crain and Vincent Gramoli and Mikel Larrea and Michel Raynal},
  title     = {{DBFT:} Efficient Leaderless {B}yzantine Consensus and its Application to Blockchains},
  booktitle = {Proc. IEEE NCA},
  pages     = {1--8},
  year      = {2018}
}

@book{Gra22,
  author    = {Vincent Gramoli},
  title     = {Blockchain Scalability and its Foundations in Distributed Systems},
  publisher = {Springer},
  year      = {2022},
}

@inproceedings{THG23,
  title     = {Smart {Redbelly} Blockchain: Reducing Congestion for {Web3}},
  booktitle = {Proc. IEEE IPDPS},
  author    = {Tennakoon, Deepal and Hua, Yiding and Gramoli, Vincent},
  year      = {2023},
  pages     = {940--950}
}

@misc{noauthor_avalanchego_nodate,
  author = {Avalanche},
  title  = {{AvalancheGo} Configs and Flags},
  url    = {https://docs.avax.network/nodes/configure/configs-flags},
  note   = {Accessed: Aug. 31, 2024},
  year   = {2024}
}

@misc{noauthor_solana_nodate,
  author = {Solana},
  title  = {{S}olana Leader Rotation},
  url    = {https://docs.solanalabs.com/consensus/leader-rotation},
  note   = {Accessed: Aug. 31, 2024},
  year   = {2024}
}

@techreport{VG20,
  title       = {Dispel: {B}yzantine {SMR} with Distributed Pipelining},
  author      = {Gauthier Voron and Vincent Gramoli},
  year        = {2020},
  number      = {1912.10367},
  institution = {arXiv}
}

@article{LSP82,
  author  = {Lamport, Leslie and Shostak, Robert and Pease, Marshall},
  title   = {The {B}yzantine Generals Problem},
  journal = {ACM Trans. Program. Lang. Syst.},
  volume  = {4},
  number  = {3},
  year    = {1982},
  pages   = {382--401}
}

@misc{algo-node-types,
  author = {Algorand},
  title  = {Algorand node types},
  url    = {https://developer.algorand.org/docs/run-a-node/setup/types/},
  note   = {Accessed: Mar. 7, 2025},
}

@misc{ava-nodes,
  author = {Avalanche},
  title  = {Nodes \& Validators},
  url    = {https://build.avax.network/docs/nodes},
  note   = {Accessed: Mar. 7, 2025},
}

@misc{apt-quorum-store,
  author = {Brian Cho and Alexander Spiegelman},
  title  = {Quorum {Store}: How Consensus Horizontally Scales on the {Aptos} Blockchain},
  url    = {https://medium.com/aptoslabs/quorum-store-how-consensus-horizontally-scales-on-the-aptos-blockchain-988866f6d5b0},
  year   = {2023},
  note   = {Accessed: Mar. 7, 2025}
}

@misc{apt-nodes,
  author = {Aptos},
  title  = {Node Networks and Sync},
  url    = {https://aptos.dev/en/network/blockchain/node-networks-sync},
  year   = {2025},
  note   = {Accessed: Mar. 7, 2025}
}

@misc{apt-reqs,
  author = {Aptos},
  title  = {Node Requirements},
  url    = {https://aptos.dev/en/network/nodes/validator-node/node-requirements},
  year   = {2025},
  note   = {Accessed: Mar. 7, 2025}
}

@misc{ava-fees,
  author = {Avalanche},
  title  = {How are gas fees calculated?},
  url    = {https://support.avax.network/en/articles/6169826-how-are-gas-fees-calculated},
  note   = {Accessed: Mar. 7, 2025},
}

@misc{ava-apr-3,
  title  = {Apricot {P}hase {T}hree: {C}-{Chain} Dynamic Fees},
  url    = {https://medium.com/avalancheavax/apricot-phase-three-c-chain-dynamic-fees-432d32d67b60},
  author = {O'Grady, Patrick},
  year   = {2021},
  note   = {Accessed: Mar. 7, 2025}
}

@misc{ava-apr-4,
  title  = {Apricot {P}hase {F}our: {Snowman}++ and Reduced {C}-{Chain} Transaction Fees},
  url    = {https://medium.com/avalancheavax/apricot-phase-four-snowman-and-reduced-c-chain-transaction-fees-1e1f67b42ecf},
  author = {O'Grady, Patrick},
  year   = {2021},
  note   = {Accessed: Mar. 7, 2025}
}

@misc{ava-block-explorer,
  author = {Avalanche},
  title  = {Blockchain explorer for {Avalanche} {L1s}},
  url    = {https://subnets.avax.network/stats/},
  note   = {Accessed: Mar. 7, 2025},
}

@misc{ava-apr-5,
  title  = {Apricot {Phase} {Five}: {P}{\textless}{\textgreater}{C} {Atomic} {Transfers}, {Atomic} {Transaction} {Batching}, and {C}-{Chain} {Fee} {Algorithm}…},
  url    = {https://medium.com/avalancheavax/apricot-phase-five-p-c-atomic-transfers-atomic-transaction-batching-and-c-chain-fee-algorithm-912507489ecd},
  author = {O'Grady, Patrick},
  year   = {2021},
  note   = {Accessed: Mar. 7, 2025}
}

@misc{ava-galaxy,
  author = {Charles Yu},
  title  = {{RL1}: Digging into {Avalanche}},
  url    = {https://www.galaxy.com/insights/research/ready-layer-one-avalanche/},
  note   = {Accessed: Mar. 7, 2025},
  year   = {2022}
}

@misc{ava-rpc-fee-cap,
  author = {Avalanche},
  title  = {C-{Chain}},
  url    = {https://docs.avax.network/nodes/chain-configs/c-chain\#rpc-tx-fee-cap},
  note   = {Accessed: Mar. 7, 2025},
}

@misc{redbelly-reqs,
  author = {Redbelly},
  title  = {Node hardware specification requirements — {Vine} ({Redbelly} developer portal)},
  url    = {https://vine.redbelly.network/nds-node-hardware-specification-requirements},
  note   = {Accessed: Mar. 7, 2025}
}

@misc{ava-fundamentals,
  author = {Michael Nadeau},
  title  = {The fundamentals of {Avalanche}},
  url    = {https://tokenterminal.com/crypto-research/avalanche\#decentralization},
  year   = {2023},
  note   = {Accessed: Mar. 7, 2025}
}

@misc{anza-turbine,
  author = {Anza},
  title  = {{T}urbine Block Propagation},
  url    = {https://docs.anza.xyz/consensus/turbine-block-propagation},
  note   = {Accessed: Mar. 7, 2025}
}

@misc{helius-turbine,
  author = {Ryan Chern},
  title  = {Turbine: Block Propagation on {Solana}},
  url    = {https://www.helius.dev/blog/turbine-block-propagation-on-solana},
  year   = {2023},
  note   = {Accessed: Mar. 7, 2025}
}

@misc{algorand-sync-issue,
  author = {Juan Granados},
  title  = {network: {Handle} failed to broadcast transactions · Issue \#5641 · algorand/go-algorand},
  url    = {https://github.com/algorand/go-algorand/issues/5641},
  note   = {Accessed: Mar. 7, 2025},
  year   = {2023}
}

@inproceedings{APLPP19,
  author    = {Emmanuelle Anceaume and Antonella Del Pozzo and Romaric Ludinard and Maria Potop{-}Butucaru and Sara {Tucci Piergiovanni}},
  title     = {Blockchain Abstract Data Type},
  booktitle = {Proc. ACM SPAA},
  pages     = {349--358},
  year      = {2019}
}

@inproceedings{vSRGG25,
  author    = {von Seck, Richard and Rezabek, Filip and Gallenm\"{u}ller, Sebastian and Carle, Georg},
  title     = {On the Impact of Network Transport Protocols on Leader-Based Consensus Communication},
  year      = {2025},
  booktitle = {Proc. ACM BSCI},
  pages     = {1--11}
}

@inproceedings{EGJ18,
  author    = {Parinya Ekparinya and Vincent Gramoli and Guillaume Jourjon},
  title     = {Impact of Man-In-The-Middle Attacks on {E}thereum},
  booktitle = {Proc. IEEE SRDS},
  pages     = {11--20},
  year      = {2018}
}

@inproceedings{HWYKS23,
  author    = {Hwanjo Heo and Seungwon Woo and Taeung Yoon and Min Suk Kang and Seungwon Shin},
  title     = {Partitioning {E}thereum without Eclipsing It},
  booktitle = {Proc. NDSS},
  year      = {2023}
}

@inproceedings{NG16,
  author    = {Christopher Natoli and Vincent Gramoli},
  title     = {The Blockchain Anomaly},
  booktitle = {Proc. IEEE NCA},
  pages     = {310--317},
  year      = {2016}
}

@inproceedings{NG17,
  author    = {Christopher Natoli and Vincent Gramoli},
  title     = {The Balance Attack or Why Forkable Blockchains are Ill-Suited for Consortium},
  booktitle = {Proc. IEEE DSN},
  pages     = {579--590},
  year      = {2017}
}

@article{NEJ24,
  author  = {Natoli, Chris and Ekparinya, Parinya and Jourjon, Guillaume and Gramoli, Vincent},
  title   = {Blockchain Double Spending with Low Mining Power and Network Delays},
  year    = {2024},
  volume  = {3},
  number  = {4},
  journal = {ACM Distrib. Ledger Technol.}
}

@article{ABDPGGVZ23,
  author  = {Karolos Antoniadis and Julien Benhaim and Antoine Desjardins and Elias Poroma and Vincent Gramoli and Rachid Guerraoui and Gauthier Voron and Igor Zablotchi},
  title   = {Leaderless consensus},
  journal = {J. Parallel Distrib. Comput.},
  volume  = {176},
  pages   = {95--113},
  year    = {2023}
}

@inproceedings{RPG24,
  author    = {Alejandro Ranchal{-}Pedrosa and Vincent Gramoli},
  title     = {{ZLB:} A Blockchain to Tolerate Colluding Majorities},
  booktitle = {Proc. IEEE DSN},
  pages     = {209--222},
  year      = {2024}
}

@inproceedings{singh_bft_2008,
  title     = {{BFT} protocols under fire},
  booktitle = {Proc. USENIX NSDI},
  author    = {Singh, Atul and Das, Tathagata and Maniatis, Petros and Druschel, Peter and Roscoe, Timothy},
  year      = {2008},
  pages     = {189--204}
}

@inproceedings{lee_turret_2014,
  title     = {Turret: A Platform for Automated Attack Finding in Unmodified Distributed System Implementations},
  booktitle = {Proc. IEEE ICDCS},
  author    = {Lee, Hyojeong and Seibert, Jeff and Hoque, Endadul and Killian, Charles and Nita-Rotaru, Cristina},
  year      = {2014},
  pages     = {660--669}
}

@InProceedings{bano_twins_2022,
  author    = {Bano, Shehar and Sonnino, Alberto and Chursin, Andrey and Perelman, Dmitri and Li, Zekun and Ching, Avery and Malkhi, Dahlia},
  title     = {Twins: {BFT} Systems Made Robust},
  booktitle = {Proc. OPODIS},
  pages     = {7:1--7:29},
  year      = {2022}
}

@inproceedings{YTB21,
  author    = {Youngseok Yang and Taesoo Kim and Byung-Gon Chun},
  title     = {Finding Consensus Bugs in {E}thereum via Multi-transaction Differential Fuzzing},
  booktitle = {Proc. USENIX OSDI},
  year      = {2021},
  pages     = {349--365}
}

@article{winter_randomized_2023,
  title   = {Randomized Testing of {Byzantine} Fault Tolerant Algorithms},
  volume  = {7},
  journal = {Proc. ACM Program. Lang.},
  author  = {Winter, Levin N. and Buse, Florena and De Graaf, Daan and Von Gleissenthall, Klaus and Kulahcioglu Ozkan, Burcu},
  year    = {2023},
  pages   = {757--788}
}

@inproceedings{ma_loki_2023,
  title     = {{LOKI}: State-Aware Fuzzing Framework for the Implementation of Blockchain Consensus Protocols},
  booktitle = {Proc. NDSS},
  author    = {Ma, Fuchen and Chen, Yuanliang and Ren, Meng and Zhou, Yuanhang and Jiang, Yu and Chen, Ting and Li, Huizhong and Sun, Jiaguang},
  year      = {2023}
}

@INPROCEEDINGS{CMZ23,
  author    = {Chen, Yuanliang and Ma, Fuchen and Zhou, Yuanhang and Jiang, Yu and Chen, Ting and Sun, Jiaguang},
  booktitle = {Proc. IEEE S\&P},
  title     = {Tyr: Finding Consensus Failure Bugs in Blockchain System with Behaviour Divergent Model},
  year      = {2023},
  pages     = {2517-2532},
}

@inproceedings{MaPhoenix23,
  author    = {Ma, Fuchen and Chen, Yuanliang and Zhou, Yuanhang and Sun, Jingxuan and Su, Zhuo and Jiang, Yu and Sun, Jiaguang and Li, Huizhong},
  title     = {Phoenix: Detect and Locate Resilience Issues in Blockchain via Context-Sensitive Chaos},
  year      = {2023},
  booktitle = {Proc. ACM CCS},
  pages     = {1182--1196}
}

@inproceedings{ZBJ25,
  title     = {Blackbox Fuzzing of Distributed Systems with Multi-Dimensional Inputs and Symmetry-Based Feedback Pruning},
  author    = {Zou, Yonghao and Bai, Jia-Ju and Jiang, Zu-Ming and  Zhao, Ming and Zhou, Diyu},
  booktitle = {Proc. NDSS},
  year      = {2025}
}

@inproceedings{AAA24,
  author    = {Mohammad Javad Amiri and Chenyuan Wu and Divyakant Agrawal and Amr El Abbadi and Boon Thau Loo and Mohammad Sadoghi},
  title     = {The Bedrock of {B}yzantine Fault Tolerance: A Unified Platform for {BFT} Protocols Analysis, Implementation, and Experimentation},
  booktitle = {Proc. USENIX NSDI},
  year      = {2024},
  pages     = {371--400}
}

@article{Zhang23,
  author  = {Zhang, Long and Ron, Javier and Baudry, Benoit and Monperrus, Martin},
  title   = {Chaos Engineering of {E}thereum Blockchain Clients},
  year    = {2023},
  volume  = {2},
  number  = {3},
  journal = {ACM Distrib. Ledger Technol. Res. Pract.}
}

@inproceedings{YQZ24,
  author    = {Yaish, Aviv and Qin, Kaihua and Zhou, Liyi and Zohar, Aviv and Gervais, Arthur},
  title     = {Speculative denial-of-service attacks in {E}thereum},
  year      = {2024},
  booktitle = {Proc. USENIX Secur.}
}

@inproceedings{LCL21,
  title     = {As Strong As Its Weakest Link: How to Break Blockchain DApps at RPC Service},
  booktitle = {Proc. NDSS},
  author    = {Li, Kai and Chen, Jiaqi and Liu, Xianghong and Tang, Yuzhe and Wang, XiaoFeng and Luo, Xiapu},
  year      = {2021}
}

@inproceedings{LWT21,
  author    = {Li, Kai and Wang, Yibo and Tang, Yuzhe},
  title     = {DETER: Denial of Ethereum Txpool sERvices},
  year      = {2021},
  booktitle = {Proc. ACM CCS},
  pages     = {1645--1667}
}

@inproceedings{FK25,
  author    = {Aviv Frenkel and Dmitry Kogan},
  title     = {An Attack on TON’s ADNL Secure Channel Protocol},
  booktitle = {Proc. IEEE S\&P},
  year      = {2025}
}

@inproceedings{TZQ25,
  author    = {Taro Tsuchiya and Liyi Zhou and Kaihua Qin and Arthur Gervais and Nicolas Christin},
  title     = {Blockchain Amplification Attack},
  booktitle = {Proc. ACM SIGMETRICS},
  year      = {2025}
}

@inproceedings{LTC21,
  author    = {Li, Kai and Tang, Yuzhe and Chen, Jiaqi and Wang, Yibo and Liu, Xianghong},
  title     = {TopoShot: uncovering Ethereum's network topology leveraging replacement transactions},
  year      = {2021},
  booktitle = {Proc. ACM IMC},
  pages     = {302--319}
}

@INPROCEEDINGS{ZZZ24,
  author    = {Zhao, Chonghe and Zhou, Yipeng and Zhang, Shengli and Wang, Taotao and Sheng, Quan Z. and Guo, Song},
  booktitle = {Proc. IEEE INFOCOM},
  title     = {DEthna: Accurate Ethereum Network Topology Discovery with Marked Transactions},
  year      = {2024},
  pages     = {1711-1720},
}

@inproceedings{HKZG15,
  author    = {Ethan Heilman and Alison Kendler and Aviv Zohar and Sharon Goldberg},
  title     = {Eclipse Attacks on Bitcoin's Peer-to-Peer Network},
  booktitle = {Proc. USENIX Secur.},
  pages     = {129--144},
  year      = {2015}
}

@techreport{WG16,
  title       = {Ethereum eclipse attacks},
  author      = {W{\"u}st, Karl and Gervais, Arthur},
  year        = {2016},
  institution = {ETH Zurich},
}

@inproceedings{SPL25,
  title     = {Eclipse attack on {M}onero’s peer to peer network},
  author    = {Shi, Ruisheng and Peng, Zhiyuan and Lan, Lina and Ge, Yulian and Liu, Peng and Wang, Qin and Wang, Juan},
  booktitle = {Proc. NDSS},
  year      = {2025}
}

@inproceedings{AZV17,
  title     = {Hijacking bitcoin: Routing attacks on cryptocurrencies},
  author    = {Apostolaki, Maria and Zohar, Aviv and Vanbever, Laurent},
  booktitle = {Proc. IEEE S\&P},
  pages     = {375--392},
  year      = {2017}
}

@inproceedings{STM18,
  title     = {POSTER: deterring {DD}o{S} attacks on blockchain-based cryptocurrencies through mempool optimization},
  author    = {Saad, Muhammad and Thai, My T and Mohaisen, Aziz},
  booktitle = {Proc. ACM ASIACCS},
  pages     = {809--811},
  year      = {2018}
}

@ARTICLE{AFO21,
  author  = {Aponte-Novoa, Fredy Andres and Orozco, Ana Lucila Sandoval and Villanueva-Polanco, Ricardo and Wightman, Pedro},
  journal = {IEEE Access},
  title   = {The 51\% Attack on Blockchains: A Mining Behavior Study},
  year    = {2021},
  volume  = {9},
  pages   = {140549-140564},
}

@article{ES18,
  title   = {Majority is not enough: Bitcoin mining is vulnerable},
  author  = {Eyal, Ittay and Sirer, Emin G{\"u}n},
  journal = {Commun. ACM},
  volume  = {61},
  number  = {7},
  pages   = {95--102},
  year    = {2018}
}

@misc{Ros12,
  author = {Meni Rosenfeld},
  title  = {Analysis of hashrate-based double-spending},
  year   = {2012}
}

@INPROCEEDINGS{NTT21,
  author    = {Neu, Joachim and Tas, Ertem Nusret and Tse, David},
  booktitle = {Proc. IEEE S\&P},
  title     = {Ebb-and-Flow Protocols: A Resolution of the Availability-Finality Dilemma},
  year      = {2021},
  pages     = {446-465},
}

@inproceedings{MJP20,
  author    = {Mirkin, Michael and Ji, Yan and Pang, Jonathan and Klages-Mundt, Ariah and Eyal, Ittay and Juels, Ari},
  title     = {BDoS: Blockchain Denial-of-Service},
  year      = {2020},
  booktitle = {Proc. ACM CCS},
  pages     = {601--619}
}

@ARTICLE{CBR22,
  author  = {Chaganti, Rajasekhar and Boppana, Rajendra V. and Ravi, Vinayakumar and Munir, Kashif and Almutairi, Mubarak and Rustam, Furqan and Lee, Ernesto and Ashraf, Imran},
  journal = {IEEE Access},
  title   = {A Comprehensive Review of Denial of Service Attacks in Blockchain Ecosystem and Open Challenges},
  year    = {2022},
  volume  = {10},
}

@ARTICLE{IST21,
  author  = {Imtiaz, Muhammad Anas and Starobinski, David and Trachtenberg, Ari and Younis, Nabeel},
  journal = {IEEE Trans. Netw. Service Manag.},
  title   = {Churn in the Bitcoin Network},
  year    = {2021},
  volume  = {18},
  number  = {2},
  pages   = {1598-1615},
}

@ARTICLE{MMM20,
  author  = {Motlagh, Saeideh G. and Mišić, Jelena and Mišić, Vojislav B.},
  journal = {IEEE Trans. Netw. Sci. Eng.},
  title   = {Impact of Node Churn in the Bitcoin Network},
  year    = {2020},
  volume  = {7},
  number  = {3},
  pages   = {2104-2113},
}

@inproceedings{DPBF25,
  author    = {Di Perna, Vincenzo and Bernardo, Marco and Fabris, Francesco and Amaro, Sebastiao and Matos, Miguel and Schiavoni, Valerio},
  title     = {Impact of Network Topologies on Blockchain Performance},
  booktitle = {Proc. ACM DEBS},
  year      = {2025}
}

@inproceedings{DPSF25,
  author    = {Di Perna, Vincenzo and Schiavoni, Valerio and Fabris, Francesco and Bernardo, Marco},
  title     = {Blockchain Energy Consumption: Unveiling the Impact of Network Topologies},
  booktitle = {Proc. IEEE ICBC},
  year      = {2025}
}

@article{DLS88,
  author  = {Dwork, Cynthia and Lynch, Nancy and Stockmeyer, Larry},
  title   = {Consensus in the presence of partial synchrony},
  year    = {1988},
  volume  = {35},
  number  = {2},
  journal = {J. ACM},
  pages   = {288–323}
}

@misc{But17,
  author = {Vitalik Buterin},
  title  = {Response to the Balance Attack},
  url    = {https://www.reddit.com/r/ethereum/comments/5rcm4o/comment/dd6b4je/},
  year   = {2017},
  note   = {Accessed: Jun. 2, 2025}
}

@inproceedings{GSF24,
  author    = {Giuliari, Giacomo and Sonnino, Alberto and Frei, Marc and Streun, Fabio and Kokoris-Kogias, Lefteris and Perrig, Adrian},
  title     = {An Empirical Study of Consensus Protocols’ DoS Resilience},
  year      = {2024},
  booktitle = {Proc. ACM ASIACCS},
  pages     = {1345–1360}
}

@misc{Red20,
  author = {Jamie Redman},
  title  = {Bitcoin Gold 51\% Attacked - Network Loses \$70,000 in Double Spends},
  url    = {https://news.bitcoin.com/bitcoin-gold-51-attacked-network-loses-70000-in-double-spends/},
  year   = {2020},
  note   = {Accessed: Aug. 26, 2025}
}

@misc{Zim20,
  author = {Terence Zimwara},
  title  = {\$5.6 Million Double Spent: ETC Team Finally Acknowledges the 51\% Attack on Network},
  url    = {https://news.bitcoin.com/5-6-million-stolen-as-etc-team-finally-acknowledge-the-51-attack-on-network/},
  year   = {2020},
  note   = {Accessed: Aug. 26, 2025}
}

@misc{Guar14,
  title  = {Hacker makes \$84k hijacking Bitcoin mining pool},
  note   = {\url{https://www.theguardian.com/technology/2014/aug/07/hacker-bitcoin-mining-pool-internet-service-providers-canada-dell}},
  year   = {2014},
  author = {Tom Brewster}
}

\end{document}